\documentstyle[12pt,aasms4,psfig]{article}
\slugcomment{To appear in the Astronomical Journal}

\lefthead{Rigopoulou et al.}
\righthead{A large mid-infrared spectroscopic and near-IR imaging survey of
ULIRGs: their nature and evolution}
\begin{document}

\title{A large mid-infrared spectroscopic and near-IR imaging survey of
ULIRGs: their nature and evolution \footnote{Based on observations 
with ISO, an ESA project with instruments funded
by ESA member states (especially the PI countries: France, Germany, the
Netherlands, and the United Kingdom) with the participation of ISAS and NASA.}
\footnote {Based on observation collected at the European Southern Observatory,
Chile, ESO No 62.P-0315.}
}
\author{D. Rigopoulou\altaffilmark{1}, H.W.W. Spoon\altaffilmark{1,2},
R. Genzel\altaffilmark{1}, D. Lutz\altaffilmark{1}, 
A.F.M. Moorwood\altaffilmark{2}, Q.D. Tran\altaffilmark{1}}

\altaffiltext{1}{Max-Planck-Institut f\"ur extraterrestrische Physik,
Postbox 1603, D-85740, Garching, Germany}

\altaffiltext{2}{European Southern Observatory, Karl-Schwarzschild-Strasse 2,
85748 Garching, Germany}

\begin{abstract}

We present a low resolution mid-infrared spectroscopic survey of an unbiased 
sample of 62
Ultraluminous Infrared Galaxies (ULIRGs) (L$_{IR} >$ 10$^{12}$ L$_{\odot}$,
z$\leq$0.3)
using ISOPHOT-S on board the Infrared Space Observatory (ISO). 
For comparison we also present ISOPHOT-S spectra for 23 active galactic nuclei
(AGN) and 15 starburst and normal galaxies. 
The  line-to-continuum ratio of the 7.7 $\mu$m PAH emission feature is used as a discriminator between starburst and AGN activity in
ULIRGs. We find that the majority of ULIRGs are predominantly powered by
starbursts. The ratio of PAH over infrared luminosities, 
L$_{\rm PAH} /$ L$_{\rm IR}$, for starburst dominated ULIRGs is very similar
to the ratio found for template starbursts.
The shapes of the PAH features are sometimes unusual. 
Extinction has a noticeable effect on the PAH spectra of ULIRG starbursts.

We have obtained high resolution near-infrared imaging for the Southern 
ISOPHOT-S ULIRGs to investigate their evolution stage.
The majority (68\%) of the ULIRGs imaged are
double systems and all show distorted morphologies.
From the 23 double nuclei systems 17 of them appear at linear separations
between 4-14 kpc with a mean separation of 6.5 kpc. 
Using the separations measured from our new near-infrared imaging as well as
from the literature,
we have examined whether ULIRGs that are advanced mergers are more AGN-like.
We have found no such evidence contrary to what is postulated by the classical 
evolutionary scenario. No correlation is found between the stage of merger in
ULIRGs and their infrared luminosity. In fact we find that systems in the 
early stages of merging may well output maximum luminosity. 
We also find that the total mass of interstellar gas, as estimated from the CO
(1$\rightarrow$0) luminosity, does not decrease with decreasing merger
separation.
When both an AGN and a starburst occur concurrently in ULIRGs, 
we find that the starburst dominates the luminosity output.
We propose that the available
gas reservoir and the individual structure of the interacting
galaxies plays a major role in the evolution of the system.
\end{abstract}

\section{Introduction} \label{intro}

With bolometric luminosities and space densities comparable to those of 
quasars (Soifer et al. 1987), ULIRGs are 
the dominant population of luminous galaxies in the local Universe. 
Sanders and Mirabel (1996)
have reviewed the properties of ULIRGs which can be summarized as follows: 
(1) most of the bolometric luminosity of ULIRGs is emitted at far-infrared
(12-100 $\mu$m) wavelengths, (2) all ULIRGs are invariably very dust and gas
rich systems (e.g. Rigopoulou et al. 1996a, Solomon et al. 1998), (3) most 
ULIRGs are interacting systems (e.g. Leech et al. 1994) 
with distorted
morphologies. The quasar like luminosities of ULIRGs led Sanders et al. (1988)
to propose an evolutionary scenario where most ULIRGs are predominantly powered
by dust enshrouded quasars.

For almost a decade, one subject of debate has been centered on the 
issue of whether the dominant luminosity source of ULIRGs is an AGN or a
starburst. The ULIRG spectrum from the radio to infrared
displays characteristics similar to those of starbursts  
(Condon et al. 1991, Crawford et al. 1996, Rowan-Robinson and Crawford 1989, 
Rigopoulou et al. 1996b, etc), and the powerful ``starburst winds''
detected in some of them (Armus et al. 1990, Heckman et al. 1990) are usually
associated with starbursts.
However, optical spectra
resembling those of Seyfert galaxies (Sanders et al. 1988, Kim et al. 1995), 
and hard X-rays, a signpost of AGN
activity (Ogasaka et al. 1997, Kii et al. 1997), have been detected in some
ULIRGs.
In reality, probably both starburst and AGN occur simultaneously in ULIRGs.

Along with the plethora of observational data on ULIRGs, it has become all the
more evident that interactions and mergers are at the heart of the ULIRG
phenomenon. Dynamical modeling has demonstrated that galaxy interactions and
mergers are efficient means of driving central inflows of gas (Noguchi et al.
1991, Barnes and Hernquist 1996, Mihos and Hernquist 1996). 

Optical imaging (Leech et al. 1994, Clements et al. 1996) has
shown that virtually all ULIRGs are interacting however, dust obscuration has
not allowed a full investigation of the interactions since some of the nuclei
are deeply embedded. That dust really plagues the appearance of ULIRGs in 
the optical has also been demonstrated through recent WFPC2 imaging (Borne et
al. 1998). In order to study the interaction and merging process in ULIRGs 
high resolution imaging at longer wavelengths, such as the near-infrared (NIR)
regime, is necessary. With the advances in detector technology
high resolution imaging in the NIR is now possible allowing one to penetrate 
through the obscuring dust. 
So far, high resolution NIR imaging has been carried out for few individual 
objects (Carico et al. 1990, Majewski et al. 1993, Armus et al. 1995) 
while Murphy et al. (1996) and Duc Mirabel and Maza (1997) presented NIR 
images for somewhat larger samples but with resolution inadequate to study
in detail the various morphological features.

With the advent of the Infrared Space Observatory (ISO) of the European Space 
Agency (Kessler et al. 1996) mid-infrared spectroscopy became available for a
large sample of ULIRGs (Genzel et al. 1998), enabling reconsideration of 
the energetics and the evolution of ULIRGs.
Observations of mid-infrared fine structure lines with the Short Wavelength
Spectrometer (SWS)
as well as of
polycyclic aromatic hydrocarbon (PAH) features found in emission at 6.2, 
7.7, 8.6 and 11.3 $\mu$m with the ISOPHOT-S for a sample of 15 ULIRGs (Genzel et
al. 1998), show that these are powerful tools in probing the luminosity source
in ULIRGs. Ground based observations of PAH features in a variety of
environments have shown that PAH are in principle strong in 
starburst galaxies but weak or absent in classical AGN (Moorwood 1986, 
Roche et al. 1991).
The observations of Genzel et al. (1998) showed that on average ULIRGs are 
starburst powered. 
To put these initial results in a more secure statistical footing the
observations were expanded to yet a larger sample of 62 ULIRGs observed with the
ISOPHOT-S, focusing on the spectral regime from 5 to 11 $\mu$m, where most of
the PAH features arise. First results from this survey have been summarized in
Lutz et al. (1998) where it was shown that 80\% of ULIRGs display starburst
characteristics with only a 20\% showing signs of AGN activity.

In this paper we present the full atlas of ISOPHOT-S spectra for 62 ULIRGs. As
discussed in Genzel et al. (1998) and Lutz et al. (1998) the PAH feature
line-to-continuum is a robust tool to discriminate between starburst 
and AGN activity in ULIRGs. For comparison we present ISOPHOT-S spectra for
template AGN and starbursts. In addition,
to gain more insight into the peculiar morphology of the 
ISOPHOT-S ULIRGs we have obtained high resolution NIR images for a 
subsample of Southern ULIRGs. 
We combine the ISO results with our high resolution imaging and probe the
evolution of ULIRGs.
The structure of the paper is as follows: the ISOPHOT-S sample is presented in
section (2) followed by the Observations and Data Reduction section (3) for 
both the ISO measurements and the NIR imaging. In Section 4 we present our
results from the ISOPHOT-S spectroscopic survey for ULIRGs, AGN and starbursts.
In section (5) the morphology of ULIRGs is presented together with a
discussion on the importance of tails as evolution indicators.
A simple
observational test of the evolution of ULIRGs together with a discussion on
merging modeling is presented in section (6) while
in section (7) we discuss possible evolutionary scenarios.

\section{ULIRG Sample} \label {sam}

The ISOPHOT-S sample (hereafter ISO-ULIRG sample) has been drawn from the 
complete 
IRAS 2 Jy (Strauss et al. 1992) and its extension IRAS 1.2 Jy sample (Fisher et
al. 1995). The selection of ULIRGs was based on the following criteria:
\begin{itemize}
\item{
luminosity L$_{FIR}
(40-120) > $ 10$^{11.7}$ L$_{\odot}$ (which corresponds to 
L$_{IR} (8-1000) \sim$ 10$^{12}$ L$_{\odot}$)\footnote{See e.g.
Sanders \& Mirabel (1996) for the definition of the 40-120 $\mu$m L$_{FIR}$ and
8-1000 $\mu$m L$_{IR}$ luminosities}}
\item{flux S$_{60} > $1.3 Jy}
\item{good ISO visibility}
\item{redshift below 0.3. The constraint on redshift was imposed to ensure that
the 7.7 $\mu$m PAH feature did not shift into regions of poor sensitivity in the
ISOPHOT-S detector.}
\end{itemize}
No color criteria have been applied
to the sample so as to not bias it towards AGN. 
The sample includes some of the bright BGS ULIRGs (Sanders et al. 1988) that
meet our selection criteria. In addition we have included 
IRAS 23060$+$0505 (S$_{60} \sim $ 1 Jy) and NGC 6240 
(less luminous however, it exhibits ULIRG characteristics). 
Accurate positions for the ISO-ULIRG sample were taken from VLA observations 
(Crawford et al., 1996), optical images (Clements et al. 1996), or from the
Automated Plate Measuring Machine (APM) using a likelihood technique. Objects
with inaccurate positions have been dropped from our ISO sample as well as
objects with poor ISO visibility. A few objects were not observed due to
the end of the ISO mission. All these effects do not pose any bias in our 
sample. Preliminary results of the ISOPHOT-S sample 
have been presented in Genzel et al. (1998),  (for most of the BGS members) 
and Lutz et al. (1998).  
Throughout the paper we have used H$_{0}$ = 75 km s$^{-1}$ Mpc $^{-1}$ and 
q$_{0}$ = 0.5.

In addition to the ISO-ULIRG sample described above, 4-16 $\mu$m spectra have 
been obtained with ISOCAM (as part of the ZZULIRG project) for a further 16
ULIRGs (hereafter ISO-ZZULIRG). This sample specifically 
populates the higher luminosity bins (L $>$ 10$^{12.25}$ L$_{\odot}$). 
A detailed analysis of the ZZULIRG sample will be presented elsewhere (Tran
et al. 1999). 

Imaging data were collected for 27 out of the 43 ISO-ULIRG plus ISO-ZZULIRG 
galaxies with 
declinations $\delta \leq$ 3$^{o}$ accessible to telescopes in the 
Southern Hemisphere. We have not imaged ISO-ULIRGs for which previous high
quality imaging data existed. Also we avoided those ISO-ULIRGs with already
available NICMOS data.

\section{Observations and Data Reduction}

\subsection{PHOT-S observations}

We obtained low
resolution ($\lambda$/$\Delta\lambda\sim$90) 2.5--11.6 $\micron$ spectra
for the entire ISO-ULIRG sample using the ISOPHOT-S spectrophotometer
(Lemke et al. 1996) onboard ISO (Kessler et al. 1996).
In addition to the ULIRG spectra we have obtained ISOPHOT-S spectra for 23 AGN
and 14 starburst and normal galaxies to be used as comparison templates.
ISOPHOT-S comprises two low-resolution grating spectrometers covering
simultaneously the wavelength range 2.47 to 4.87$\micron$ and 5.84 to
11.62$\micron$. The spectrum is registered by two linear arrays of 64
Si:Ga detectors with a common entrance aperture of 24$''$ $\times$ 24$''$.
For the observations described here we only make use of the
long-wavelength section, because the signal-to-noise ratio (S/N) for
all but the brightest members of the ISO-ULIRG sample is very low in
the short-wavelength regime. The measurements were carried out in rectangular
chopped mode, using a chopper throw of 180$''$. The resulting spectra
are thus free of contributions from zodiacal light, that would otherwise
seriously affect the faint ULIRG spectra. The pure on-source integration times
vary between 512 and 2048 s for ULIRGs and 128 -- 2048 s for AGN and starbursts.

The ISOPHOT-S data were reduced using PIA version 7.2.\footnote {PIA is  a
joint development by the ESA Astrophysics Division and the ISOPHOT
Consortium.}
The data reduction steps can be summarized as follows:\\
1) deglitching on ramp level \\ 
2) subdivision of ramps in sections of 32 non destructive
read-outs (NDRs), which greatly increases the effectiveness of
statistical tools (to be carried out in step 4) 
and takes care of minor ramp non-linearity effects. 
For faint targets such as the ULIRGs, the gain in statistics prevails \\
3) ramp fitting to derive signals \\ 
4) masking of bad signals by eye-inspection\\
5) kappa sigma and min$/$max clipping on remaining signal distribution\\
6) determination of average signal per chopper plateau\\
7) masking or correction of bad plateaus by eye-inspection\\
8) background subtraction using all but the first two plateaus, and finally,\\
9) flux calibration, using the default PIA version 7.2 spectral response 
function (SRF).

Since ISOPHOT-S was operating close to its sensitivity limits, the
detector response to changes of illumination consists of an immediate
jump to an intermediate level, followed by an extremely slow transient
to the final signal level. In practice, only the initial step is
observed. To correct for this effect, we applied a flux correction
factor of 1.4 (U. Klaas 1997, private communication) to all our chopped
ISOPHOT-S spectra.

We have compared the PIA version 7.2 default SRF with the signal dependent
SRF of Acosta-Pulido (1999) for the case of faint ULIRGs. The comparison showed
that the overall flux correction factor (1.4) we applied to our spectra
calibrated with the default SRF brings these results 
close to those obtained by using the signal dependent SRF without
correction factor. Only towards the longest PHT-SL wavelength
pixels does our overall correction factor tend to be too large.
For the PHT-SS range the situation is quite different. Here the
signal dependent SRF not only drastically changes
the continuum slope, it also removes the deep 3.1$\micron$ pseudo-``ice''
feature present in most of our standard calibrated AGN and starburst
comparison spectra. Since this feature always appears at the same
{\it observed} instead of {\it rest} wavelength it follows that 
the feature cannot be real.
We conclude that the calibration for the useful part of our ISOPHOT-S
spectra (the 5.8--11$\micron$ range) is good, and moreover,
internally consistent.

\subsection{Near-Infrared Imaging}

We have obtained high quality NIR images for 27 out of the 43 ISO galaxies with
declinations $\delta \leq$ 3$^{o}$ accessible to telescopes in the
southern hemisphere. 
Near-infrared imaging for the 27 southern ISO-ULIRG plus ISO-ZZULIRG  
galaxies was carried out in
November 1998 with the SOFI (Moorwood, Cuby and Lidman 1998) near-infrared 
imaging camera, installed on the
ESO-NTT telescope, LaSilla. For the observations we used the J band 
and K$_{s}$ band filters with
central wavelengths 1.247 $\mu$m and 2.162 $\mu$m, respectively.
The SOFI detector is a Hawaii HgCdTe 1024$\times$1024 array. 
For the current observations the projected scale of the
array is 0$^{\prime\prime}$.144 pixel$^{-1}$. Sky conditions were 
photometric throughout the acquisition of the images.
For each galaxy observations of a nearby star were made in order to obtain an
accurate estimate of the point spread function (PSF) at the time of the
observation. Integration time per frame
was set to 20 sec, reaching a total integration time of 20 mins for the K$_{s}$
band (except for 01355-1814 for which integration time in K$_{s}$ was 8 mins) 
and 6 mins for the J band.

The observations were carried out using the ``jitter'' technique which is the
most efficient method of sky acquisition with a minimum loss of observing time.
In this technique the telescope is offset around the central position (the
offsets were generated randomly but were restricted within a 20$^{\prime\prime}$
square box 
around the central position) by small amounts. A typical acquisition in jitter
mode consisted of 6 frames for the J- band and 20 frames for the K$_{\rm s}$
band. Sky estimation and subtraction, flat-fielding (based on dome-flats),
plane-recentering, and frame-coaddition were all performed using the ``jitter''
algorithm which is part of the ECLIPSE software collection designed especially
for the reduction of SOFI and ISAAC images (Devillard 1997).

The photometry has been carried out in IRAF using circular apertures centered on
each nucleus. For flux calibration , faint stars from the NICMOS, UKIRT Faint
Standards lists have been observed before and after the galaxy
observations.
Under the excellent seeing conditions (FWHM range between 
0.45$^{\prime\prime}$ -- 0.8$^{\prime\prime}$)
we estimated that the errors
in the reported magnitudes are less than 0.08 mag. 
Figure 1 presents the calibrated K$_{\rm s}$ contour maps of the
27 galaxies that we have observed (since for the current discussion we only use
the K$_{\rm s}$ band data we only present those maps. The J band data will be
discussed in a forthcoming paper). The linear scale is indicated.

\section{ISOPHOT-S Results}

\subsection {PAH features}

The 3-12 $\mu$m window in the spectra of most galaxies contains a 
number of broad emission features, the most important being at 3.3, 6.2,
7.7, 8.7 and 11.2 $\mu$m. These features appear whenever the 
interstellar medium is exposed to moderately intense
UV radiation (see e.g. review in Puget \& L\'eger 1989). The carriers of these
features are either large carbon-rich
molecules, the so-called Polycyclic Aromatic Hydrocarbons (PAH), or very small
grains consisting of amorphous aromatic carbons (e.g. Sakata et al. 1987).
Although the precise nature of the carriers is still somewhat uncertain, we 
will refer to them as ``PAH'' which is one of the most popular identifications, 
noting that their use as a diagnostic tool for external galaxies is unaffected 
by  this uncertainty. Since their first discovery as ``unidentified''
features (Gillett et al. 1973), the mid-infrared features
have been detected in a variety of Galactic environments. ISO observations
are adding substantial detail to PAH feature studies of galactic regions
relevant to extragalactic studies, such as PDRs, HII
regions (e.g. Verstraete et al. 1996, Roelfsema et al. 1996), and 
the diffuse emission of the galactic disk (Mattila et al. 1996). 

Prior to ISO, studies of PAH features in galaxies were limited. Following their
discovery in the spectrum of M82 (Gillett et al. 1975, Willner et al. 1977),
several studies noted that starbursts exhibit more
prominent PAH features than Seyferts, in the latter the emission is usually 
either weak
or absent (e.g. Moorwood 1986, Roche et al. 1991). With ISO we can now 
study the 
PAH emission in a large, unbiased sample of ultraluminous galaxies, which
we have supplemented by starburst and AGN comparison objects.

\subsection{ISOPHOT-S Spectroscopy}

The spectra for the 62 ULIRGs, the template AGN and starbursts
observed with ISOPHOT-S are shown in Figures 2, 3 and 4 respectively.
Among the AGN templates
there are several known composite sources with both AGN and starburst activity.
For instance, NGC 7469 is a Seyfert 1 nucleus surrounded by a star forming ring
which contributes up to two thirds of the bolometric luminosity (Genzel et al.
1995). A starburst ring is also contributing 
to the bolometric luminosity of
Circinus (Moorwood et al. 1996), the radio galaxy Cen A, and the narrow line
X-ray galaxy NGC 7582. 
Our ISOPHOT-S data for the template AGN and starbursts confirm the findings 
from previous 
observations (e.g. Roche et al. 1991): we find that the 7.7 $\mu$m emission 
feature is strong in dusty 
environments with moderately strong UV radiation fields, such as those found 
in starbursts. PAH emission becomes significantly weaker 
in very strong radiation fields such as those found in AGN. 
Overall, the strength of the PAH line to continuum ratio in AGN is about an 
order of magnitude weaker than the template starbursts. The strength of the
7.7$\mu$m emission feature in the above mentioned composite galaxies ranges
between these two extremes. 

The majority of the ULIRG ISOPHOT-S spectra show PAH features at
6.2, 7.7 and 8.6 $\mu$m similar to those seen in starburst galaxies. 
This is also evident in their average spectrum, presented in
Figure 1 of Lutz et al. (1998). Given that ULIRGs may host starburst and AGN,
we assume the PHT-S spectra to be a superposition of a PAH dominated starburst
spectrum and a continuum dominated AGN spectrum, both likely modified by 
significant obscuration. To estimate the relative contributions, we use 
here a simple empirical method  also employed by Genzel et al. (1998) 
and Lutz et al. (1998), based on the line-to-continuum ratio of the
strongest feature (7.7$\mu$m). We show in Figure 2 the continua 
we have adopted in this process.

Continuum determination is important and not
trivial, since the 5-11 $\mu$m range is characterized by a multitude of 
features. The PAH emission features overlap with each other and with
the broad silicate absorption centered at 9.6 $\mu$m, and this overlap is likely 
to be important for obscured galaxies like ULIRGs. We have determined the 
continuum by linear
interpolation between two pivot points at 5.9 $\mu$m
and 10.9 $\mu$m (both rest wavelength). For objects whose redshift
moves the 10.9 $\mu$m point out of the observed range we used the empirical
formula
S$_{10.9}$ = (2.5 $\pm$ 0.5) $\times$ S $_{5.9}$ which we found adequate for
low redshift sources. Continuum determination by eye
was done for a handful of template AGN where the method described above gave 
unphysical
results (for instance interpolated 7 $\mu$m continuum above the observed
spectrum). The
peak strength of the 7.7$\mu$m feature was measured by simply taking
the average of all data within a window covering the rest wavelengths
7.57--7.94 $\mu$m. Finally, the ratio of feature minus (local) continuum and
continuum was determined.  Errors or limits on all these quantities are
based on the noise measured shortward of 5.9 $\mu$m rest wavelength and
the increase in noise towards longer wavelengths corresponding to the
decrease of the ISOPHOT-S spectral response.

The selected 5.9 $\mu$m point is largely free of feature emission and 
provides a secure
point for the continuum determination. On the other hand, the 10.9 $\mu$m
point is less securely determined since it is still within the long
wavelength side of the silicate
absorption feature. Therefore, one can assume that our adopted continuum
is a ``lever'' which can be rotated significantly around the 5.9 $\mu$m point.
The placement of the 10.9 $\mu$m continuum level is not arbitrary and the
slope of the continuum cannot be very steep. There is a physical reason for
this: the 10.9 $\mu$m point should be such that the resulting
continuum should never rise above the data in the 6.5--7.7 $\mu$m range since 
the PAH features are known to sit on a plateau of the same origin.
Although the uncertainty in placement of the continuum can affect the 
strength of the 7.7 $\mu$m PAH L$/$C, it will not turn a ``strong PAH'' into a
``weak or absent PAH'' and therefore our method in separating
starbursts from AGN based on PAH features remains robust.

Low resolution spectra of the 6.2--11.3 $\mu$m PAH features and the
underlying plateau are usually well fit by Lorentzian or Cauchy
functions (e.g. Boulanger et al. 1998 for PDRs, Mattila et al. 1999
for the galaxy NGC 891). For the ULIRGs we do not adopt such a fitting
procedure to keep our method unbiased towards a ``canonical'' PAH shape
which will not be adequate for low feature-to-continuum ratio systems
like Mrk 231 whose weak feature, be it of PAH or different origin, is
not fit well by a standard PAH shape.  Similar results to those
presented in this work are obtained using two other methods to
quantify the relative importance of PAH emission (Tran et al. 1999).
In the first one, the importance of PAH features is determined from a
simple ratio of fluxes measured in two narrow bands at 7.7 and
5.9 $\mu$m. The other method implements a more elaborate fit of the data,
using variable contributions of an AGN continuum and a starburst PAH
spectrum, both affected by (different) extinction.

The feature-to-continuum ratio of the 7.7$\mu$m PAH feature is
presented in Tables 1, 2 and 3 for the ISOPHOT-S ULIRG sample, the AGN
and the starburst templates, respectively.  Along with other
information we quote the flux densities at 5.9 and 10.9$\mu$m, the
points used for continuum determination.  Table 4 gives values for the
properties of the {\it averaged} ULIRG, starburst and AGN spectrum
(see Lutz et al. 1998).  The strength of the 7.7 PAH L$/$C is used to
classify ULIRGs as starbursts or AGN. All ULIRGs with a 7.7 PAH L$/$C
$\leq$1 (or upper limits slightly above 1) are classified as AGN and
the rest as starbursts.

\subsection{L$_{\it \bf{PAH}}/$L$_{\it \bf{IR}}$: are starburst dominated 
ULIRGs similar to starbursts ? }

In Figure 5 we compare the properties of ULIRG-starbursts (those
ULIRGs with PAH L$/$C $>$ 1) to those of the template starbursts (all
starbursts from Table 3 except M81, M51 and NGC 4569).  In each case
we compute the ratio of the 7.7$\mu$m PAH luminosity, L$_{\rm PAH}$, over
far-infrared L$_{\rm IR}$ luminosities.  As can be seen from Figure 5
the distribution of the L$_{\rm PAH} /$ L$_{\rm IR}$ is found to be
similar in ULIRG-starburst and template starbursts.  The mean L$_{\rm
PAH} /$ L$_{\rm IR}$ value for starbursts is 8.13$\times$10$^{-3}$,
only 1.4 times larger than the value found for ULIRG-starbursts. The
L$_{\rm PAH} /$ L$_{\rm IR}$ ratio provides an estimate of the
``activity'' of the galaxy in the sense that the more starbursting
activity a galaxy shows the smaller the ratio becomes due to the higher
infrared luminosity (e.g. Acosta-Pulido et al. 1996).

Mattila et al. (1999) show that the 5.9--11.3 $\mu$m PAH luminosity in
NGC 891, a normal non-starburst system, makes up 9\% of the total IR
luminosity of the system. Allowing for the difference between 7.7
$\mu$m luminosity and ``total'' PAH emission (5.9--11.3 $\mu$m) we
find that in a normal galaxy the PAH emission contributes a higher
fraction to the total IR radiation than in a starburst galaxy. Moving
the comparison to ULIRG-starbursts and starbursts we see exactly the
same trend: ULIRG-starbursts show a somewhat lower L$_{\rm PAH} /$
L$_{\rm IR}$ ratio 
(log[{\rm mean value}] = -2.26) than the template starbursts
(log[{\rm mean value}] = -2.09).  Nonetheless, the comparison of the 
L$_{\rm PAH}/$L$_{\rm IR}$ ratios implies that ULIRG-starbursts have similar
properties to template starbursts and that our method to separate
ULIRG-starbursts from ULIRG-AGN is correct.

\subsection{The effect of extinction on ULIRG PAH spectra}

A close inspection of Figure 2 suggests that the observed ratios of
the PAH features in ULIRGs differ somewhat from those in lower luminosity
starbursts which exhibit rather homogeneous PAH properties.
For only a few individual ULIRGs like Arp 220 is the S/N of the ISOPHOT-S
spectra sufficient to show 
this effect, but the average ULIRG spectrum (Lutz et al. 1998)
confirms its overall relevance.  Compared to the 7.7 $\mu$m feature, which is 
the strongest feature in ULIRGs as well as in starbursts, 
the 6.2 $\mu$m and 11.3 $\mu$m 
(where available) features are somewhat weaker in the ULIRGs. The same 
applies to the 
8.6 $\mu$m feature which is sometimes just a shoulder of the 7.7 $\mu$m 
feature rather than a well-separated feature (see e.g. the case of Arp 220). 

The weakness of 6.2 and the 8.6 $\mu$m features may reflect intrinsic
variations related to the unusual conditions in the interstellar
medium in ULIRGs.  There is evidence for such variations from ISO
observations of HII regions and reflection nebulae in our own Galaxy
(e.g.  Verstraete et al.  1996, Roelfsema et al. 1996, 1998, Cesarsky
et al. 1996).  Various explanations including changes in the PAH
species mix, PAH dehydrogenization, and PAH ionization have been put
forward by these authors.  Some of the H II region spectra of
Roelfsema et al. (1998) resemble the average ULIRG spectrum, due to
intrinsic variations or due to extinction, as suggested below for the
ULIRGs. While most PAH spectra of our comparison starbursts (Fig. 4)
are remarkably similar, we believe that a real intrinsic variation is
present in at least one source, the low metallicity starburst NGC
5253.  Similarly, the faint residual (PAH?) features in the spectrum
of NGC 1068 do not fit a normal PAH spectrum.

We believe, however, that the dominant influence on the ULIRG PAH
ratios is extinction. This has already been suggested by Lutz et
al. (1998) on the basis of an overall consistency between average
ULIRG extinction derived from SWS spectroscopy, (extinction-induced)
weakening of the 6$\mu$m continuum, and PAH ratio variations. In
Figure 6 we illustrate the effect of extinction by comparing the Arp
220 PHOT-S spectrum with the M82 spectrum for which additional screen
extinction of A$_{\small V}$$\sim$20 and A$_{\small V}$$\sim$50 has
been applied. The M82 spectrum is the ISO-SWS spectrum
(F\"orster-Schreiber 1999) degraded to PHT-S resolution.  In such an
obscured starburst spectrum, the 11.3 $\mu$m feature is weak, the 8.6
$\mu$m feature has become a shoulder of the stronger 7.7 $\mu$m
feature, and the 6.2 $\mu$m feature has been suppressed relative to
the 7.7 $\mu$m one, similar to what we observe for the ULIRGs. The
assumed screen extinction is, of course, a simplification, and
uncertainties in the extinction curve cannot be ignored.
Nevertheless, these feature changes due to obscuration reproduce well
the features of Arp 220, considering that the 11.3$\mu$m feature of
Arp 220 is on top of a stronger continuum. This steeply rising
continuum is difficult to locate at the end of the PHOT-S spectrum,
but seen more clearly in the data of Charmandaris (1997) and Smith et
al. (1989). Most plausibly, it reflects the extreme concentration of
star forming activity in the inner few 100 parsecs of Arp220, leading
to a stronger ``warm'' continuum due to HII regions in terms of the
three component model of Laurent et al. (2000).
                                                                               
The effect of extinction on the 6.2$/$7.7 $\mu$m PAH ratio is demonstrated in 
a more quantitative way in Figure 7. Here,
the 6.2$/$7.7 PAH flux ratio is plotted as a
function of the extinction A$_{\small V}$ as estimated from 
independent ISO-SWS spectroscopy
(from Genzel et al. 1998, values converted to screen case)\footnote{We assume
that the obscuration is due to a foreground screen of dust with optical depth 
$\it \tau(\it \lambda) = 0.916 A (\it \lambda)$}. 
To estimate the extinction A$_{\small V}$ we used 
the extinction curve
of Draine (1989) together with the extinction curve derived from ISO--SWS 
data on the Galactic center (Lutz et al. 1996).
The total 6.2$\mu$m
and 7.7$\mu$m feature fluxes have been derived by simple integration of the
continuum-subtracted spectrum over the rest wavelength ranges 6.0-6.5 and 
7.3-8.2$\mu$m, respectively. 
The diagram includes data for template starburst galaxies,
the molecular ring in the Galactic Center, and for those ULIRGs for which
extinction estimates are available from SWS spectroscopy. 
Figure 7 shows that there is a clear
anticorrelation between the 6.2$/$7.7 PAH ratio and extinction. Of the
templates, NGC 4945 and the molecular ring in the Galactic Center have
extinctions that approach those of ULIRGs. Interestingly, those are the only
templates that show low 6.2$/$7.7 PAH ratios and their 8.6
$\mu$m features appear as a shoulder of the 7.7$\mu$m feature rather than as a
separate feature, due to suppression in the wings of the silicate absorption
feature.
How robust is the 6.2$/$7.7 $\mu$m ratio to the location of the underlying
continuum in our method of measuring feature fluxes? 
We have used the Arp 220 spectrum as a test case and varied the 10.9
$\mu$m continuum point between $\sim$70 and $\sim$400 mJy. Within this range
the 6.2$/$7.7 $\mu$m feature ratio changes between 0.18 to 0.20 but stays well
below the typical lower extinction starburst values (Figure 7).
Thus our conclusion is robust to plausible variations in the continuum.

With extinction being important, could the observed ULIRG PHT-S
spectra in fact be explained by pure absorption of a smooth underlying
continuum?  The ISO spectra of heavily obscured young stellar objects
(e.g. d'Hendecourt et al. 1996, Whittet et al. 1996) show a ``pseudo
emission feature'' created by a decrease in absorption between the
broad 9.6$\mu$m silicate feature and the shorter wavelength 6.8 and
6.0 $\mu$m absorption features. With limited wavelength coverage, this
could be misinterpreted as PAH emission. This is likely not the case
for the ULIRG PHOT-S spectra, however.  Firstly, the ULIRGS have a
clear 6.2 $\mu$m emission feature. In an absorption spectrum, such as
the above mentioned young stellar objects, another ``pseudo-emission''
can arise between the 6.0 and 6.8 $\mu$m features, but its peak will be
at about 6.5 $\mu$m (d'Hendecourt et al. 1996, Whittet et al. 1996),
incompatible with the ULIRG observations.  Secondly, the short
wavelength rise of the 7.7 $\mu$m feature in the ULIRG spectra is very
steep. This is typical for PAH emission, but not easily reproduced
with an absorption spectrum. An ambiguity exists for the feature at 8
$\mu$m. Self-absorbed silicate emission from an AGN torus (e.g. Pier
and Krolik 1992) could mimic the PAH feature here.  While its role in
some of the more AGN-like ULIRGs remains to be investigated, we find
this explanation implausible for the ULIRGs given the steep short
wavelength rise of the 7.7 $\mu$m feature in both the average ULIRG
spectrum and individual starburst-like ULIRGs like Arp 220.

Since extinction is affecting the strength of the PAH features, with
the 6.2, 8.8 and 11.3 $\mu$m being suppressed more than the 
7.7 $\mu$m,
it is rather uncertain to derive the strength of the PAH features
based only on the 11.3 $\mu$m strength alone as Dudley (1999) has
done, in particular if the PAHs are assumed to be unobscured. Such a
PAH determination is likely to underestimate the real strengths of the
features.  Any attempts to model the strength of PAHs should include
extinction effects.

\section{ULIRGs: interacting$/$merging galaxies}

\subsection{Infrared Morphologies}

That mergers and interactions play an important role in the formation
and evolution of ULIRGs is now well established (e.g. Sanders and
Mirabel, 1996).  In this section we focus on the morphology of
individual targets and aim to understand the transition through the
various stages of the merging process, as might be delineated through
our near-infrared imaging.  Nuclear separations of all the sample
galaxies with apparent double nuclei have been measured: angular and
linear separations respectively are shown in columns 5 and 6 of Table
5 for our ISO-ULIRG sample, and columns 2 and 3 of Table 6 for the
ULIRGs whose separations were taken from the literature or from the
NICMOS archive.
In order to classify a ULIRG as a double nucleus system, both nuclei must be
clearly visible and spatially distinct in the K$_{\rm s}$-band images. 
Almost all double nuclei are at same redshift as determined by
our followup near-infrared spectroscopy (Rigopoulou et al. 1999) or
published optical spectroscopy (Duc Maza and Mirabel 1997, Kim et al. 1998)

It is obvious (Figure 1) that the vast majority of the ISO-ULIRGs appear to be
double, interacting, merging systems. Of the 27 objects we imaged and the 
additional 6 NICMOS objects, 68\% appear to have double nuclei, with
a wide range of projected separations between the nuclear components.
Moreover, features such as tidal tails, bridges connecting nuclei, double
(or even multiple)
nuclear peaks stress the connection between enhanced far-infrared emission and
galactic interactions (e.g. Joseph \& Wright 1984, Cutri \& McAlary 1985 etc).
We discuss the importance of tails in section 5.2.

Only two of the single-nuclei ULIRGs, 
01003--2238 and 05189--2524, appear to be star-like with no sign of 
recent merging activity. Most single nuclei show clear signs
of interactions: 23578--5307, 00406--3127, 00183--7111 
01199--2307, 20049-7210, 23529-2119, 00397-1312 all show fairly extended 
structure.

Among the close double$/$interacting nuclei, there appears to be a variety
of morphological differences. 
The nuclear separations as a function of the recessional velocities for the
ISO-ULIRG sample as well as ULIRGs with additional literature data are 
plotted in Figure 8. 
The single nuclei are plotted with upper limits on their nuclear separations.
Separations down to 1$^{\prime\prime}$ 
are detected out to large recessional velocities
although these velocity bins above 80000 km$/$s are underpopulated in the
initial ULIRG sample. However, this does not introduce any bias in the
classification scheme or in the nuclear separations statistics. 

Double nuclei systems vary from close nuclei like 03521$+$0028,
06009-7716 and 04063--3236, to more distant interacting members like
01166--0844, 04114--5117 and finally to systems where the interaction
involves more than two nuclei.  Representative members of the the
latter category include 02411$+$0354 (at least three nuclei visible in
the NIR) 06206-6315 and systems like 23389--6139 and 0019--7426.  The
closest measured projected nuclear separation is that of 23230--6926
(1.03 kpc) while the most widely separated pair is 00188--0856
(projected linear separation 14.11 kpc) which may well be a
characteristic example of a still well separated system, probably in
the early stages of the merging process.

The case of chance superpositions, ie the possibility that some of the
double nuclei objects are simple superpositions of field galaxies,
could affect the classification of the double nuclei systems.
However, as Murphy et al. (1996) point out the chance that a field
galaxy will overlap with a particular ULIRG is smaller than 1\%. 
Projection effects
could in principle affect the classification of ULIRGs.  However, as
we discuss in more details in section 6, deprojected nuclear
separations are larger than the projected ones and are only affecting
the relative distance between the two nuclei but not the single vs
double nucleus classification.

Figure 9 shows a histogram of the separations of the various
ISO-ULIRGs. We find that among the 23 ULIRGs with measured
separations, 17 of them are found at separations between 4-14 kpc with
a mean separation of about 6.5 kpc. Among the 23 double nuclei systems
18 of them have measured PAH L$/$C (for the rest of them the S$/$N is
too low to make an accurate determination of the continuum of the 7.7
$\mu$m feature).  Out of those 18 ISO-ULIRGs, 14 of them have a PAH
L$/$C $>$ 1 implying starburst dominating luminosity. Among the
remaining 15 single (or unresolved double) nuclei ULIRGs, 13 of them
have measurable PAH L$/$C and among those only 6 of them have PAH
L$/$C $<$ 1. These simple statistical findings imply that there is no
clear trend for the single nuclei ULIRGs to show more AGN-like
characteristics. This issue is discussed in more detail in section 6.

\subsection{Tidal Tails and Mergers}

Toomre and Toomre (1972, hereafter T\&T) were among the first to model the 
creation of tidal
tails and bridges in galaxies. Further numerical simulations (e.g. Barnes and 
Hernquist 1996) established that tidal tails can be used to assess the stage 
of the interaction. Tidal tails can therefore be considered as a useful 
``clock of the interaction''.

We use tails and bridges to probe the interaction$/$merger status,
adopting the following classification. According to our scheme, ULIRGs
can be classified as follows: (a) fully relaxed systems, where we see
only a single point-like nucleus with relatively little or no tails,
(b) systems where the merger is completed, in such systems we see a
single nucleus but with significant residual structure or tail
formation (c) interacting pairs where the interacting nuclei are
clearly visible and can be found in a variety of separations.
The
classification of the ISO-ULIRGs according to the presence of tails is
shown in Table 5.  Although the classification of ULIRGs is somewhat
subjective ( for a proper classification dynamical evidence is also needed), 
our intention in the following discussion is to stress the
importance of the detailed morphological structure (ie tails) in
assessing the stage of the interaction. For each category we discuss a
few selected examples.

How many galaxies do we find in each of the above mentioned
categories?  Starting from the quiescent stage, the fully relaxed
systems (category a) we note that we find only two such systems,
05189-2524 and 01003-2524. In the next stage, completed mergers (category b),
there is quite a diversity. There are systems like 00406-3127,
01298-0744, 23529-2119, 00397-1312 and 00183-7111 which show distorted
morphologies with some tails, although not as pronounced. It is very
probable that these systems have completed the merger and are now
slowly evolving into a more relaxed system. In the same category (b)
we have classified 23253-5415, 02455-2220 and 23578-5307, all three
systems show pronounced tails implying a rather recent completion of
the merging process. The giant arm coming off the S side of
23523-5415 implies that the interaction has just been completed or
that the projection is not favorable (there is no sign of a second
nucleus). Similarly, 02455-2220 appears to be a single nucleus object
with a long tail running NE of the galaxy's main body.
Finally, 23578-5307 displays similar characteristics. Again, the main
(merged) galaxy shows a disturbed morphology, while an arm-like
structure extends to the E, SE of the main body of the galaxy.  Most
probably all these systems have just completed the merger phase,
however, the relics of the interaction are still prominent.

However, it is not always that single nucleus objects (ie advanced
mergers) display tails. Many of the galaxies which we classified as
interacting pairs, (category c) show extended structure, tails
and$/$or bridges (see T\&T for a more detailed discussion on formation
of tails and bridges). 
06206-6315 is a spectacular example of a violent interaction
where two (if not three) nuclei are involved, and where there is a
hint of an arm or tidal tail to the NE of the two nuclei. Clearly,
this system is still in the close interaction phase, with the two
nuclei still clearly visible.  What is even more interesting is that
east of the main interacting nuclei, there is a formation of some
kind, perhaps material being torn apart from one of the two main
nuclei.  A similar case is encountered in 04063-3236. A closely
interacting system with a hint of tidal arms running NE-SW of the
interacting body. The arm is not as clearly delineated as in
06206-6315. These two interacting galaxies probably have had a close
encounter and they are now orbiting each other.
Finally, 23389-6139 and 0019-6139 are examples of systems displaying bridges
among the interacting components. It is very likely than in both cases
the interaction involves multiple components although the exact stage
of the interaction (early or advanced) cannot be inferred easily.

Although it is hard to draw firm statistical conclusions about the
interaction stage by looking through the various tidal features that
the galaxies display, we can, at least quantify the percentage of
ULIRGs at each stage.  The results are displayed in Table 7. A small
percentage of ULIRGs , $\sim$7\%, are fully relaxed systems with no
sings of recent interaction. A somewhat larger fraction 22\%, has
completed the merging process and is probably moving towards a fully
relaxed system, still showing however, prominent tails but no second
nucleus is visible. 50\% or more of ULIRGs are still interacting,
since both nuclei can be seen in the images.
We can, therefore, use the morphological characteristics displayed by ULIRGs
to probe their evolutionary stage.

\section{Testing the Evolutionary Scenario for ULIRGs}

The QSO equivalent luminosities of ULIRGs led Sanders et al. (1988) to
propose an evolutionary scenario in which ULIRGs are the precursors of
QSOs. According to this plausible scenario interactions and merging of
the ULIRG parent sample cause gas to be transported to the inner parts
of the galaxies.  This central gas concentration triggers powerful
starburst activity.  As the merger advances the starburst activity
subsides and the gas that accretes onto the nucleus proper creates a
new or fuels an existing AGN whose luminosity increases rapidly and
dominates the bolometric output during the ULIRG phase.  Eventually,
all the obscuring dust is shed and the ULIRG develops into a normal
QSO.

The obvious prediction of this scenario is that the more advanced
mergers should, on average, be more AGN like. In other words there
should be a correlation between some indicator of merger advancement
(like projected nuclear separation) and some diagnostic that measures
the relative contribution of the AGN to the bolometric$/$IR
luminosity. Our mid-infrared spectroscopic survey and in particular
the PAH L$/$C ratio has proven to be a powerful tool (Genzel et
al. 1998, Lutz et al. 1998) in measuring the contribution of the AGN
to the ULIRG bolometric luminosity.  The mid infrared observations and
their comparison to optical spectroscopy (Lutz, Veilleux and Genzel
1999) indicate that {\em part} of the Sanders et al.  (1988) scenario
is incorrect: most ULIRGs are not buried QSOs. The second part remains
to be tested: ULIRG evolution from starburst to AGN may still exist
even if the obscured AGN phase is less important. To test the latter
part, independent indicators of evolutionary status such as nuclear
separations and merger models are required.  It is clear that high
resolution near-IR imaging data are crucial in determining accurate
separations.  A first attempt to test this scenario was made by Lutz
et al. (1998) where the data for the nuclear separations were taken
from the literature.  For this incomplete set of data for the nuclear
separations Lutz et al. found that there appears to be no trend toward
AGN dominance with decreasing separations.

We now combine our new high resolution imaging data, with NICMOS archival 
data and with data from the samples of Duc,
Mirabel and Maza (1997), Graham et al. (1990), Majewski et al. (1993) and 
Murphy et al. (1996) to obtain nuclear separations for over 50 ULIRGs. 
In Figure 10 we plot the PAH L$/$C as a function of nuclear separations. 
{\it No obvious correlation is found between the stage of the merger 
(measured by the nuclear separations) and the dominant energy source (measured
by the PAH L$/$C).}
In addition, we find that starburst dominated ULIRGs can be found over the 
entire range of observed separations from 0.3 kpc up to 15 kpc.

We use the {\em Student's t-test} to evaluate the differences between
the distribution of the separations in ULIRG-AGN and
ULIRG-starbursts. For the calculation of the standard error for the
difference of the means, {\it s$_{\rm D}$}, we have also used the
upper limits but have given them a lower weight.  We computed the
t-value
 
\begin{math}
t = \frac{x_{mean1} - x_{mean2}}{s_{D}}
\end{math}

and found it to be t=0.24. The mean distributions of the
ULIRG-starbursts and the ULIRG-AGN are therefore not statistically
different (at the 87\% confidence level).
In other words, the similarity found in the means of the two
distributions (of ULIRG-AGN and ULIRG-starburst) is real. This simple
statistical test strengthens our conclusion that there is no
correlation between the stage of the ULIRG-merger and the dominant
luminosity source (AGN or starburst).

So far, we proved that there is no correlation between stage of the
merger and the dominant luminosity using the projected nuclear
separations. However, we still need to ensure that the absence of
correlation is not a projection effect.  We mentioned earlier that the
mean projected separation of our ISO-ULIRG sample is d$_{\rm mean,
proj}$ = 6.5 kpc, which will correspond to a real separation d$_{\rm
mean, real}$ = 6.5$/$ cos$\theta$, where $\theta$ is the inclination
angle. Since we do not know the exact value of $\theta$ for each
system we can estimate the

\begin{math}
\large{
 rms(cos \theta _{\rm average}) = \int_{0}^{360} rms(cos\theta) 
 d\theta = 0.707 }
\end{math}

Therefore, the real mean separation can be as large as 10 kpc. 

Let us now take a look at typical distances and timescales throughout
the course of a merger.  For this purpose we follow the models of
Barnes and Hernquist (1996) or the later ones by Dubinski, Mihos and
Hernquist 1999 (in what follows we will refer to their model A).  What
we are interested in is how the distance between the two interacting
galaxies changes with time, as the merger proceeds through the various
interaction stages.  We note that the timing of the first event is
dependent on a number of parameters, such as for instance, how far
apart were the two galaxies when the interaction started, and the
masses of the progenitor galaxies.  After the first close encounter
(at t=0) the two interacting galaxies will start drifting apart to a
maximum distance of 50 kpc (aphelion) and then approach each other
again during the second passage. This phase lasts for about
6$\times$10$^{8}$ yrs, at which point the second close passage
occurs. From that point onwards the interaction proceeds rapidly. The
galaxies will drift apart again, reaching a maximum distance of 16 kpc
and then the third and final close passage (after t=7 $\times$10$^{8}$
yrs) happens at which point they will merge completely.  Our measured
(as well as the deprojected) separations are all smaller than $\sim$50
kpc, the distance to the first aphelion determined by model (A) of
Dubinski, Mihos and Hernquist (1999), so all of the imaged ULIRGs are
in fact closely interacting$/$merging.

As we have already mentioned, the timescales to the first close
passage, the aphelion distance and the total time to complete merger
are strongly dependent on a number of parameters such as the initial
distance between the two galaxies, their masses, the inclination of
their orbits, etc.  Although Barnes and Hernquist (1996) present a
number of models for various combinations of the parameters we note
that the phase space of most parameters is limited. For instance the
initial distance of the two interacting systems cannot be very large
because then the systems cannot become bounded.  Another crucial
parameter is the star formation rate (this will determine how quickly
the gas will be used up).  From the discussion above, we conclude that
the dominant luminosity source in ULIRGs in a given time, is most
probably determined by local effects such as the compression of the
interstellar gas, the fueling of the AGN and to a lesser extend by the
state of the merger.

\section{Merging and Evolution scenarios for ULIRGs}

The highly distorted morphologies of ULIRGs, the high frequency of
double nuclei among them together with the results from our simple
test on the classical ``evolutionary'' scenario of Sanders et al. lead
to the conclusion that a more detailed insight into the particulars of
the merger procedure is more likely to give valuable clues on the
issue of the evolution of ULIRGs and the dominant luminosity source.

In a galactic interaction, the transfer or orbital energy through
dynamical friction to internal degrees of freedom in the system marks
the onset of the merger. The angular momentum of each nucleus is
reduced significantly at each close encounter and after a few such
close passes the orbits finally become circular and the two nuclei
spiral together. The gas in each of the merging galaxies loses angular
momentum too, and concentrates towards the central region. At that
stage, the gas can be used to fuel an AGN, a starburst or
both. However, the time each galaxy spends at each of these steps is
not fixed. According to merger modeling (e.g. Mihos 2000) the result
of the various close encounters varies. For instance there is the
``pre-collision'' stage which is relatively short and the interaction
is probably mild, the ``impact'' stage with an even shorter duration
but with strong imprints of the interaction on the disk of the
galaxies (perhaps formation of tidal tales), the ``self-gravitating''
stage which leads into a more closely bound interacting system. After
the close encounter phase the final merging takes place followed by
the final relaxation of the merger.

Based on our PAH L$/$C database and our high resolution imaging what
can we say for the stage of the merging and the dominant energy
sources in ULIRGs?  The high percentage of double nuclei systems
(68\%) among ULIRGs gives the first indication that the ULIRG phase is
not confined to the latest stages of mergers. 03521+0028 (log
L$_{IR}$=12.46, d$_{linear}$ = 3.42 kpc), is an example of a high
luminosity ULIRG, displaying starburst characteristics (PAH L$/$C
3.23) where the two nuclei have not yet experienced the final
merging. However, its starburst-powered luminosity has reached maximum
levels.  In Figure 11 we have plotted the nuclear separation as a
function of luminosity for the ISO-ULIRG sample. For comparison we
have plotted data for a sample of Luminous Infrared Galaxies (LIRGs)
from Gao et al. (1996) and Gao and Solomon (1999). We have computed
the correlation coefficient (Pearson-C test) for the L$_{IR}$
vs. separation correlation. We find r = -0.15 for 30 ULIRGs (we note
that for the computation of the correlation coefficient we have not
used data for ULIRGs with upper limit values for the separations).
Clearly, this indicates that there is a very weak (not significant)
anticorrelation.  In other words the ULIRG luminosity stays, on
average, about the same at different separations. The most luminous
ULIRGs are found over a wide range of separation and {\em not} just
exclusively at the smallest separations.   
Turning to the LIRGs the computed correlation
coefficient of r=0.18 similarly implies a very weak correlation
between L$_{IR}$ and separation.  Therefore, it seems likely that the
maximum activity can be reached at any stage during the merger and not
necessarily at the final stages after the merger has been completed.
However, this is not what
would be expected on the basis of the star formation rate scaling with
the average gas density as a Schmidt law R$_{\star} \propto$ M$_{\it
gas}$ $\rho ^{\alpha} _{\it gas}$, $\alpha$ = 0.5 --1. Using this
parameterization Mihos \& Hernquist (1996) predict that most of the
star formation events take place in the final merging of two nuclei in
a single one.

Mihos and Bothun (1998) reached a similar result by studying the
H$_{\alpha}$ kinematics in four ULIRGs. In their sample they do not
see a trend for increased luminosity or, ULIRG activity, in systems
with more centrally concentrated H$_{\alpha}$. Clearly, the collisions
will drive the gas towards the center (where it can fuel an AGN or a
starburst) however, the ULIRG phase or the maximum luminosity may
happen at any given point in the merging sequence and not in the very
final stages when the merger has taken place.

To probe further the relation (if any) between molecular content and
stage of merger we have examined the correlations between separation
and observed CO luminosity and, separation and star-formation
``efficiency'' L$_{IR}/ $ M$_{H_{2}}$. In both cases we compare the
correlations for ULIRGs with those for LIRGs (CO data for ULIRGs and
LIRGs are from Gao and Solomon 1999).  In Figure 12 we have plotted
the nuclear separation as a function of L$_{IR}/ $ M$_{H_{2}}$. No
correlation is found between L$_{IR}/ $ M$_{H_{2}}$ and separation for
ULIRGs (correlation coefficient r=-0.07). However, a somewhat more
significant anticorrelation is found for LIRGs, with a correlation
coefficient r=-0.41. This would imply that the star formation rate in
LIRGs normalized to the available molecular mass, increases with
decreasing separation in agreement with the conclusion of Gao and
Solomon (1999).

There is no correlation (Figure 13) between CO luminosity, L$_{CO}$,
and separation for ULIRGs (r=-0.046). We find that for ULIRGs the
values of L$_{CO}$ remain almost constant and vary between
0.3$\times$10$^{10}$ L$_{\odot}$ and 1.6$\times$10$^{10}$ L$_{\odot}$.
Clearly, there is an evident correlation for LIRGs (r=0.61). The
correlation indicates that for LIRGs L$_{CO}$ decreases as the merger
progresses to advanced stages.  We suggest that the difference between
ULIRGs and LIRGs in the L$_{CO}$ vs. separation correlation indicates
(together with the short starburst timescales discussed in the next
paragraph) that the star formation process in major mergers is not
terminated by the systems running out of gas reservoir but by the
negative feedback of star-formation.

The timescales of the bursts in ULIRGs are estimated to be short, that
is no more than a few tens of millions of years in the most luminous
systems, and perhaps as small as 5--10$\times$10$^{6}$ years $\ll$
t$_{\it merger}$ (Genzel et al. 1998, Goldader et al. 1997, Thornley
et al. 1999).  These timescales are smaller than the estimated
duration of the ULIRG phase, found to be about 10$^{7}$--10$^{8}$
years.  How can the two different timescales be reconciled?  The
models of Barnes and Hernquist (1996) and Mihos and Hernquist (1996)
predict that strong compression of the interstellar gas happens during
or after every 2 or 3 close approaches. Strong gravitational torques
remove a large fraction of the angular momentum of the gas which
results in the gas being compressed. If star formation follows a
Schmidt law then star formation can occur during brief phases over a
few dynamical timescales (of the order of a few ten million years) at
each close approach (pericenter).  An alternative to this would be to
appeal to an AGN for the extra luminosity required.  The presence of
the AGN will shorten the difference in the timescales (duration of the
ULIRG phenomenon and burst lifetime). However, only in a few cases
does there exist evidence for the presence of an AGN contributing
significantly to (although not dominating) the bolometric output. Such
cases include for instance UGC5101 (compact IR dust source, Genzel et
al. 1998), and Mrk 273 (optical spectroscopy, X-rays, Rigopoulou et
al. et al. 1996b).

Although from the present data we cannot exclude the presence of an
AGN in ULIRGs, the AGN is definitely in a very ``quiet'' stage. Also
since the occurrence of AGN in single nuclei systems is low there is
no proof for an evolutionary scenario in which after the final stages
of the merger the system should host an AGN.  A clear example of this
is 20551-4250 which appears to be a single nucleus, probably having
just undergone merging (as evidenced by the presence of tails), yet
with a PAH L$/$C of 2.33.  Moreover, its optical spectrum (Kim et
al 1998) shows signs of starburst activity.

So what happens to the ULIRG phase?  Most likely the ULIRG phenomenon
should be attributed to the burst events which are initiated through
gas compression in every close approach of the two nuclei.  A number
of other factors must contribute to the ULIRG phenomenon as well.
Based on our high resolution imaging and input from merger models, at
this stage, we can only make a list of possible factors contributing
to the ULIRG phenomenon. These include the geometry of the encounter,
the size of the galaxies and their structures and also the available
gas in the two merging disks. As molecular measurements of ULIRGs have
shown (Rigopoulou et al. 1996a, Downes and Solomon 1998) ULIRGs are
very gas rich galaxies. The exact geometry of the encounter will
determine the number of close approaches. In turn the amount of gas
available will regulate the burst events. The individual structure of
each galaxy will determine the lifetime of the interaction until final
merger and transition to the quiescent stage. Whether or not a black
hole will be created or may already have existed in one of the two
nuclei, is probably not connected to the ULIRG phenomenon and
certainly we have no evidence of such a black hole contributing
significantly to the bolometric luminosity.

\section{Conclusions}\label{concl}

We have presented an atlas of the ISOPHOT-S spectroscopic survey of 62
ULIRGs (L$_{IR} >$ 10$^{12}$ L$_{\odot}$) plus comparison templates of
AGN, starbursts and normal galaxies, which was carried out as part of
the MPE ISO Central Program.  The range covered by the ISOPHOT-S
spectra 5-12 $\mu$m is the region of the PAH emission features. We use
the line-to-continuum ratio of the 7.7 PAH feature to distinguish
between AGN and starburst activity.

We find that, on average, ULIRGs exhibit properties resembling those
of starbursts. The ratios of PAH over infrared luminosities, L$_{\it
PAH} /$ L$_{\it IR}$, are found to be similar for starburst dominated
ULIRGs and template starbursts.  We find that extinction severely
affects the strength and shape of the PAH features. The PAH spectrum
of ULIRGs resembles very much that of an obscured starburst as is
demonstrated from a comparison of the spectrum of Arp 220 with that of
M82 for which additional extinction has been applied.

As a followup of the ISOPHOT-S ULIRG sample we have obtained high
resolution NIR imaging for a complete sample of Southern ULIRGs. The
majority of ULIRGs (68\%) appear to contain double nuclei. Among the 23
double nuclei systems 17 of them appear at linear separations between
4-14 kpc, with a mean separation of 6.5 kpc. The presence of tails and
bridges in almost all ULIRGs implies that they are undergoing
merging. We find that up to 50\% of the ULIRGs imaged have not yet
experienced the final merging.

An observational test of the Sanders evolutionary scenario shows that
there is no trend for advanced mergers to be more AGN-like. In fact we
find that the mean distribution of separations between ULIRG-AGN and
ULIRG-starburst are similar. We estimate that projection effects are
not likely to significantly affect our results. We find that AGN or
starburst dominated ULIRGs can be found at all nuclear separations.

Although an AGN and a starburst may concurrently occur in ULIRGs there is no
proof for an evolution of the type ULIRG-starburst $\rightarrow$ ULIRG-AGN. 
The dynamical lifetime of the ULIRG phase could be explained by burst
events only. No evidence is found for AGN to dominate even among the fully
merged systems. We have found no correlation between infrared luminosity output
and merger stage (or separation), a similar behavior is found in a 
sample of LIRGs studied here as well. 
 
Finally, we propose that the particulars of the merging$/$interacting
nuclei, such as the gas available in the progenitor galaxies and the
individual galaxy structure may also determine the merger procedure
and the ULIRG evolution.

\acknowledgments

We are grateful to Andreas Eckart, Volker Springel, and Linda Tacconi
for discussions and useful suggestions.
We thank the anonymous referee for his$/$her comments and suggestions.
SWS and the ISO Spectrometer Data Center at MPE 
are supported by DLR (DARA) under grants 50 QI 8610 8 and 50 QI 9402 3.
This work has made use of the NASA$/$IPAC Extragalactic Database which is
operated by the Jet Propulsion Laboratory, California Institute of technology,
under contract with the National Aeronautics and Space Administration.

\newpage

\centerline{\bf FIGURES}

\figcaption[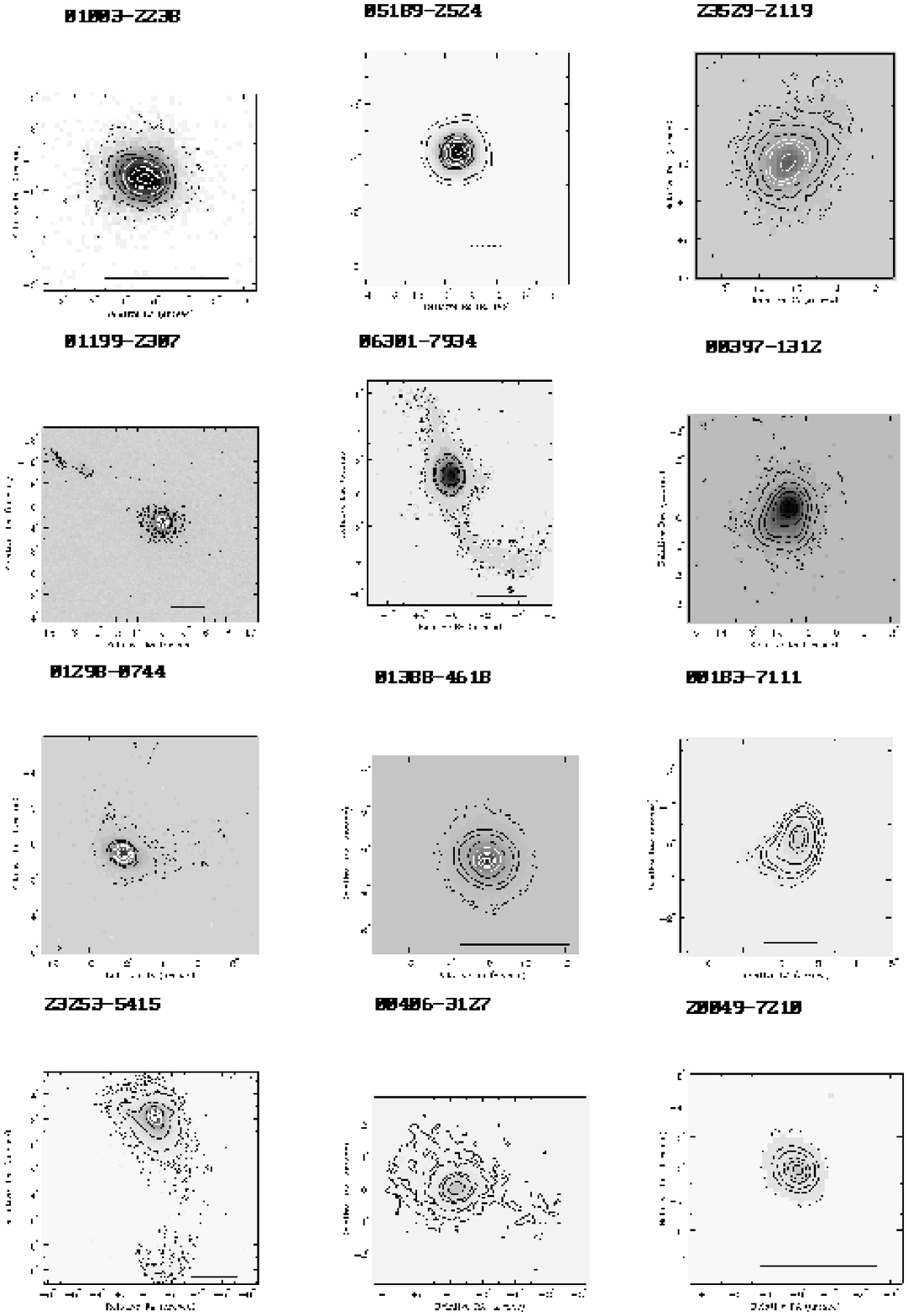, 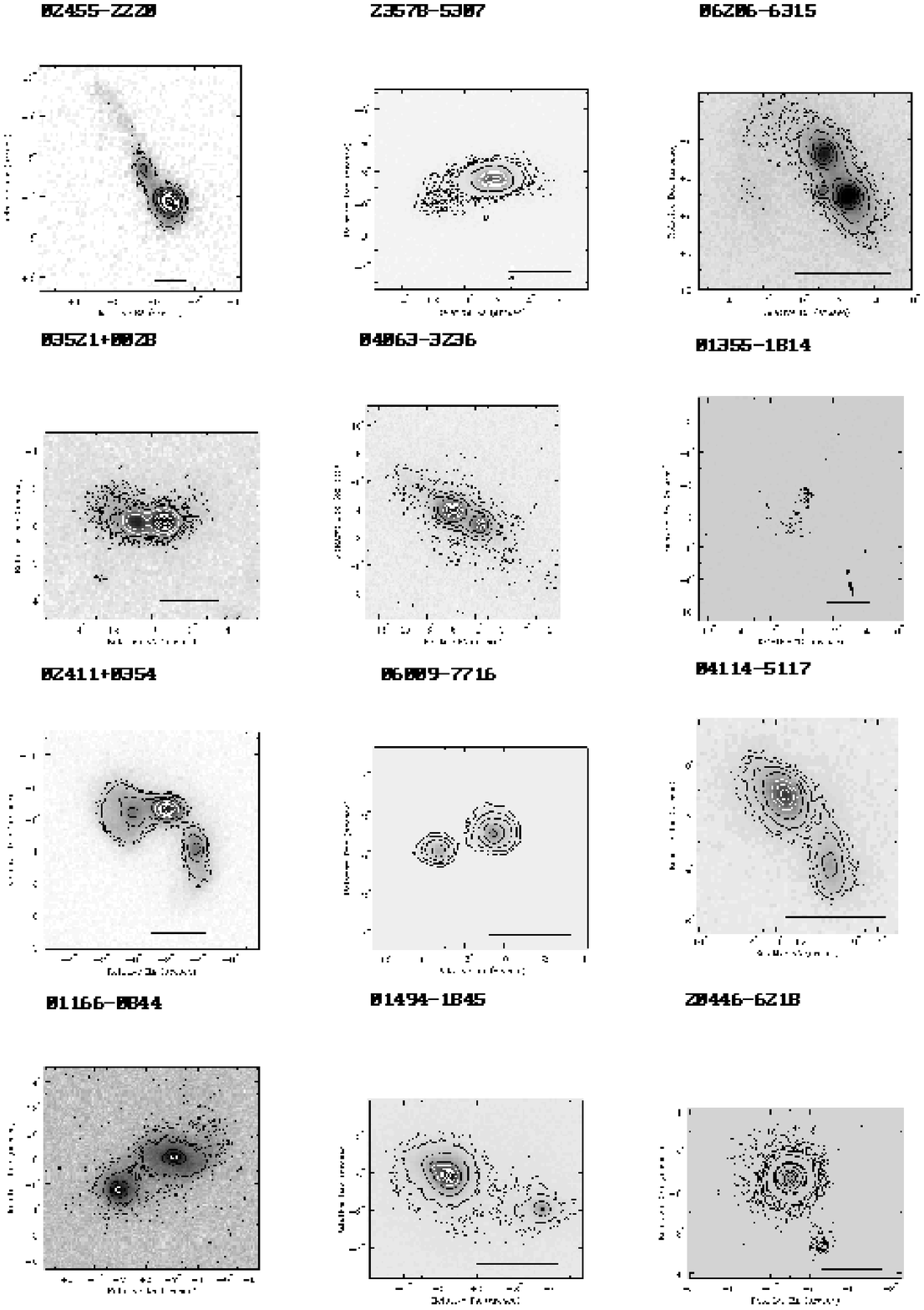, 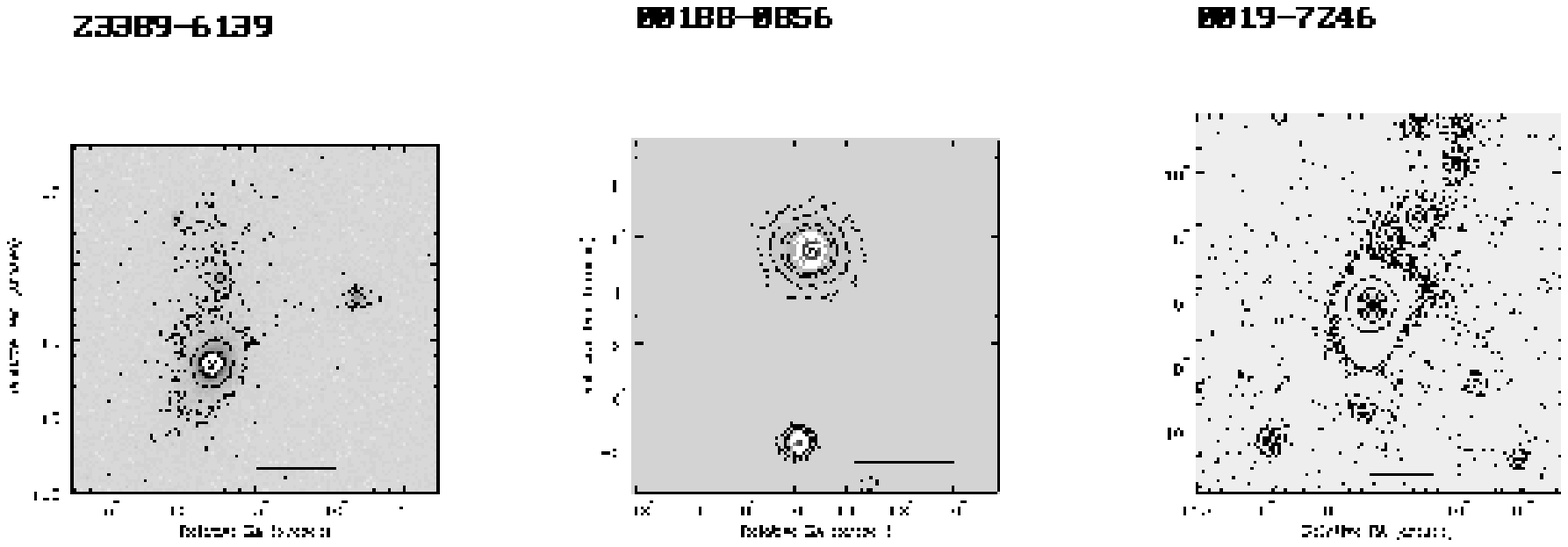]
{Near-infrared images for the 27 Southern ULIRGs presented here. 
The images are arranged according to increasing nuclear separation. The
K$_{\rm s}$ band images are plotted in grey scale and contour forms. 
The faintest contours correspond to 20.5 mag$/ \Box ''$.
Angular scale on the sky is indicated by the bar
which represents 10 kpc, except for 05189-2524 for which the bar represents
5 kpc. \label{fig:rigopoulou1}}

\figcaption[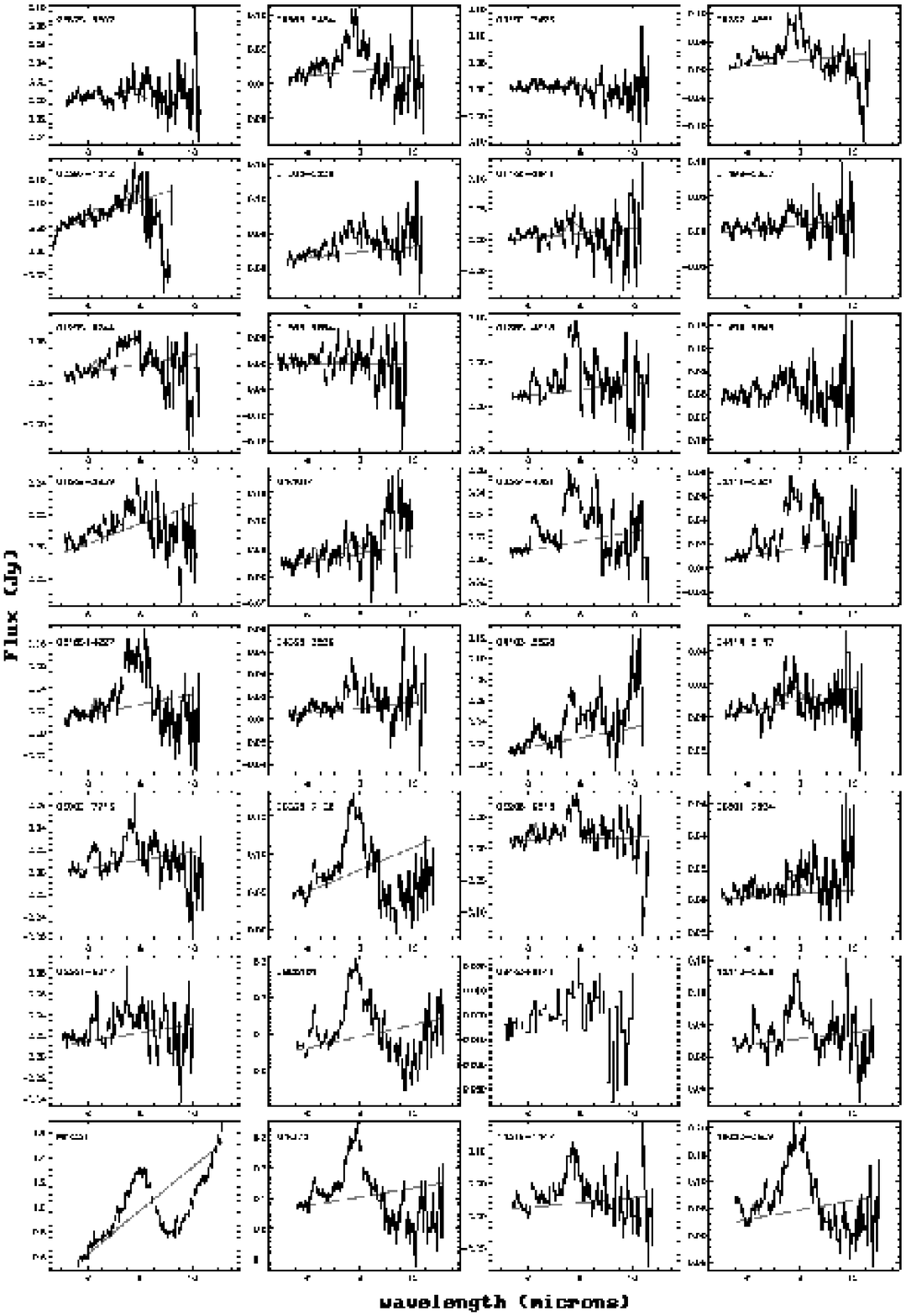,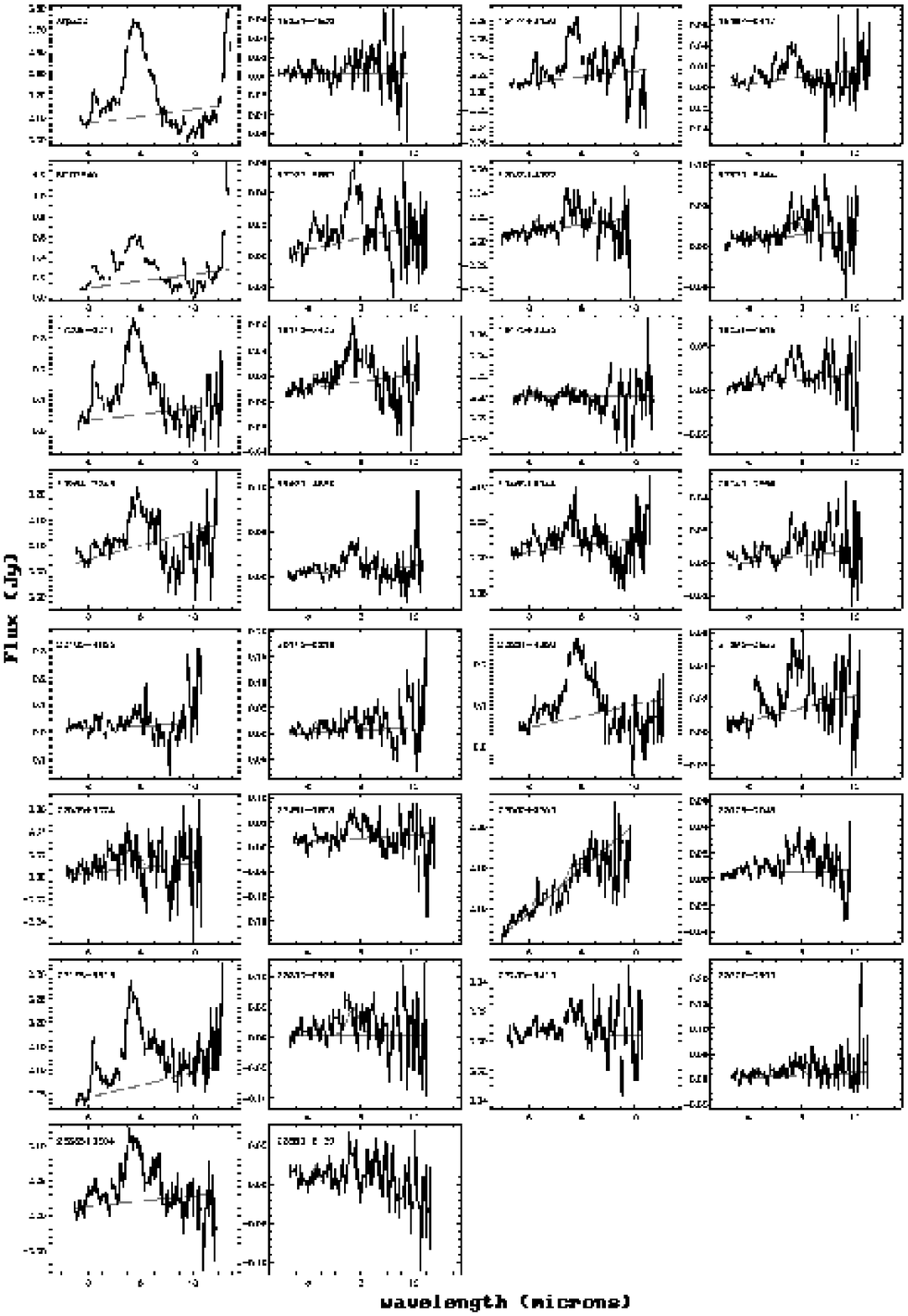]
{ISOPHOT-S spectra for the ISO-ULIRGs. The assumed continuum around
the PAH features is shown as a straight line. In case of upper limits the
Gaussian used to estimate the upper limit is shown. 
No continuum fit for 09463$+$8141 (see note 1d).
\label{fig:rigopoulou2}}

\figcaption[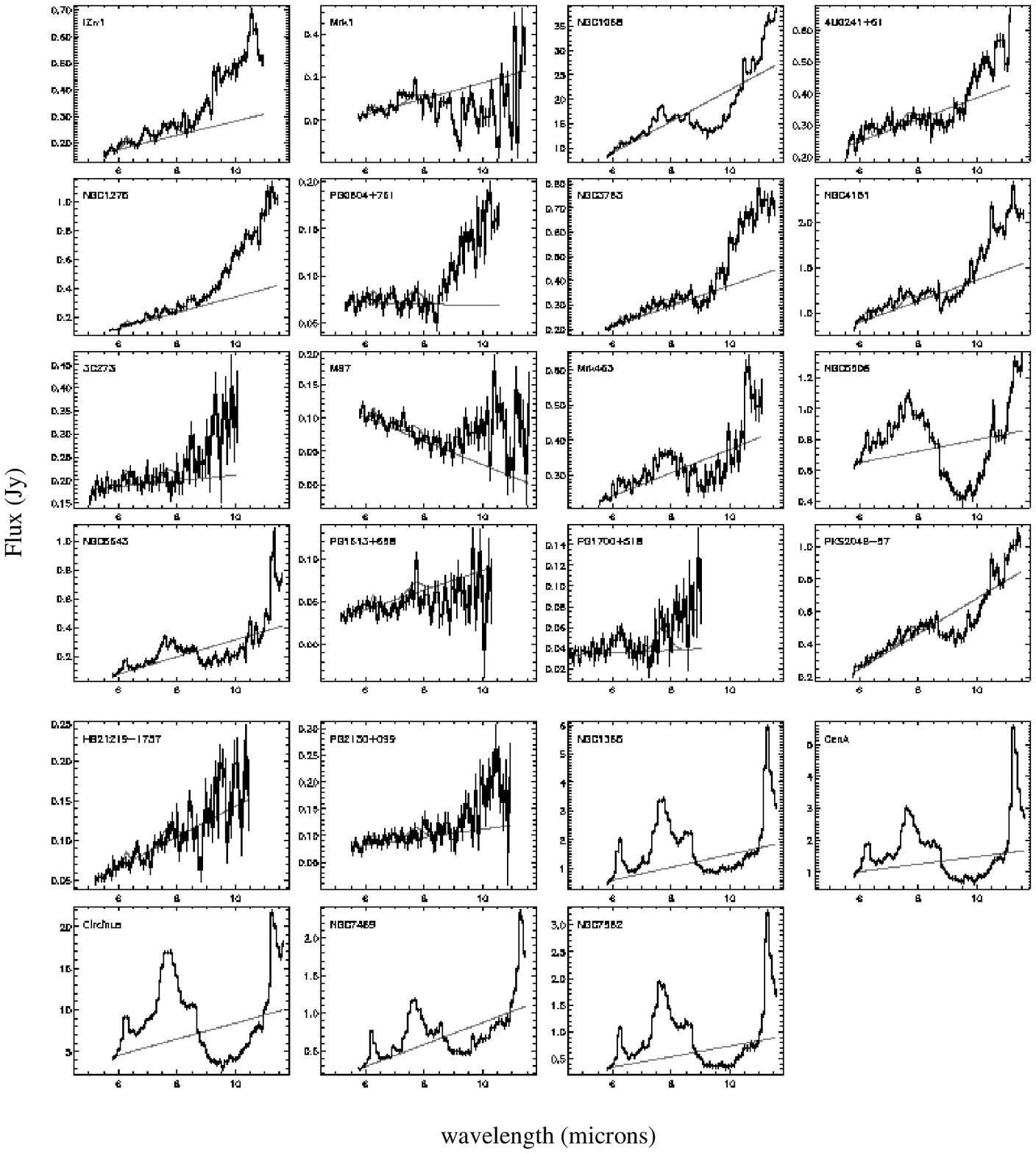]
{ISOPHOT-S spectra for the ISO-AGN (continuum and upper limits as in
Fig.2). \label{fig:rigopoulou3}}

\figcaption[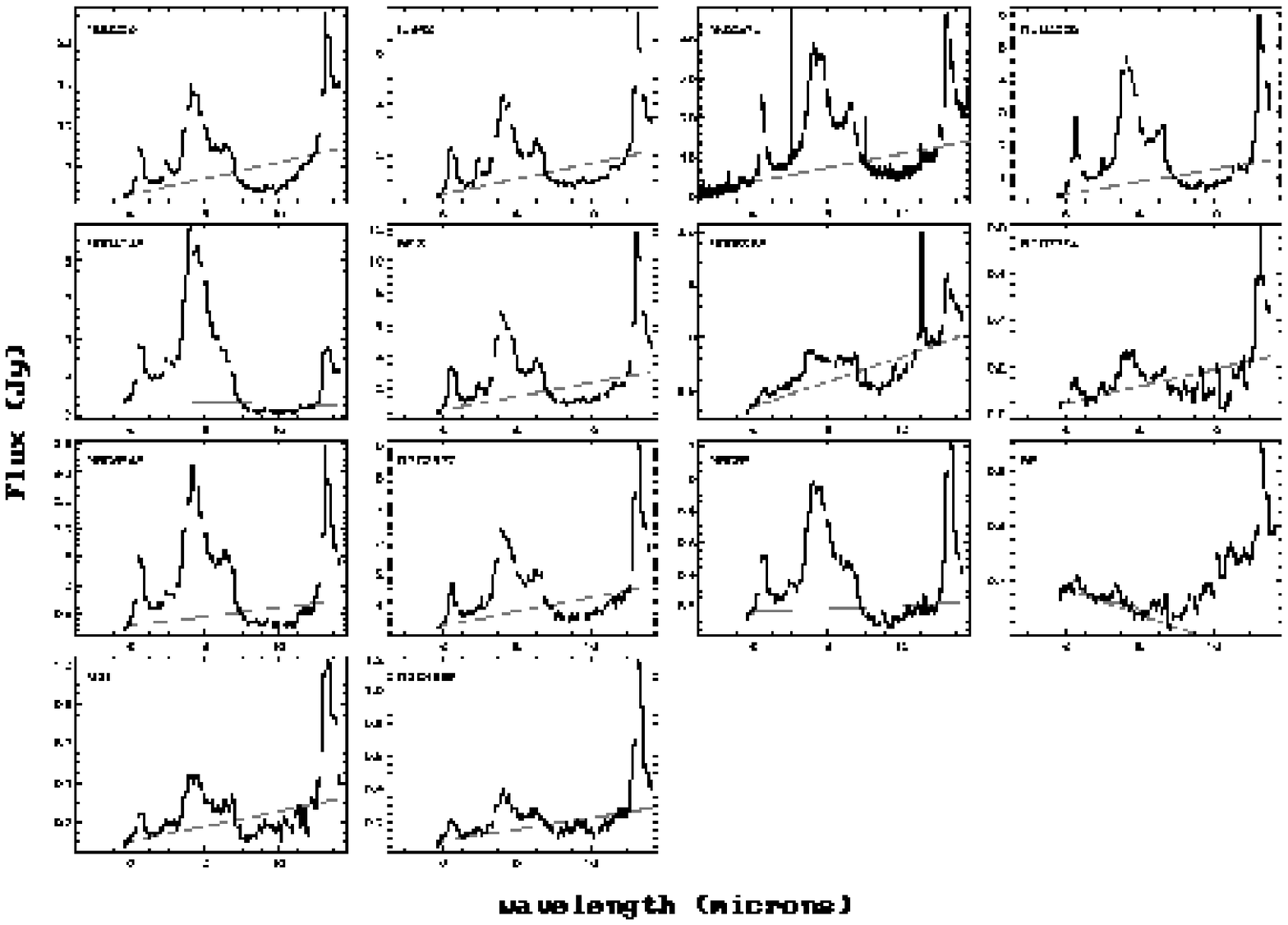]
{ISOPHOT-S spectra for the ISO-starburst, except for M82 where part of
an SWS AOT1 spectrum is presented (continuum and upper limits as in Fig. 2).
\label{fig:rigopoulou4}}

\figcaption[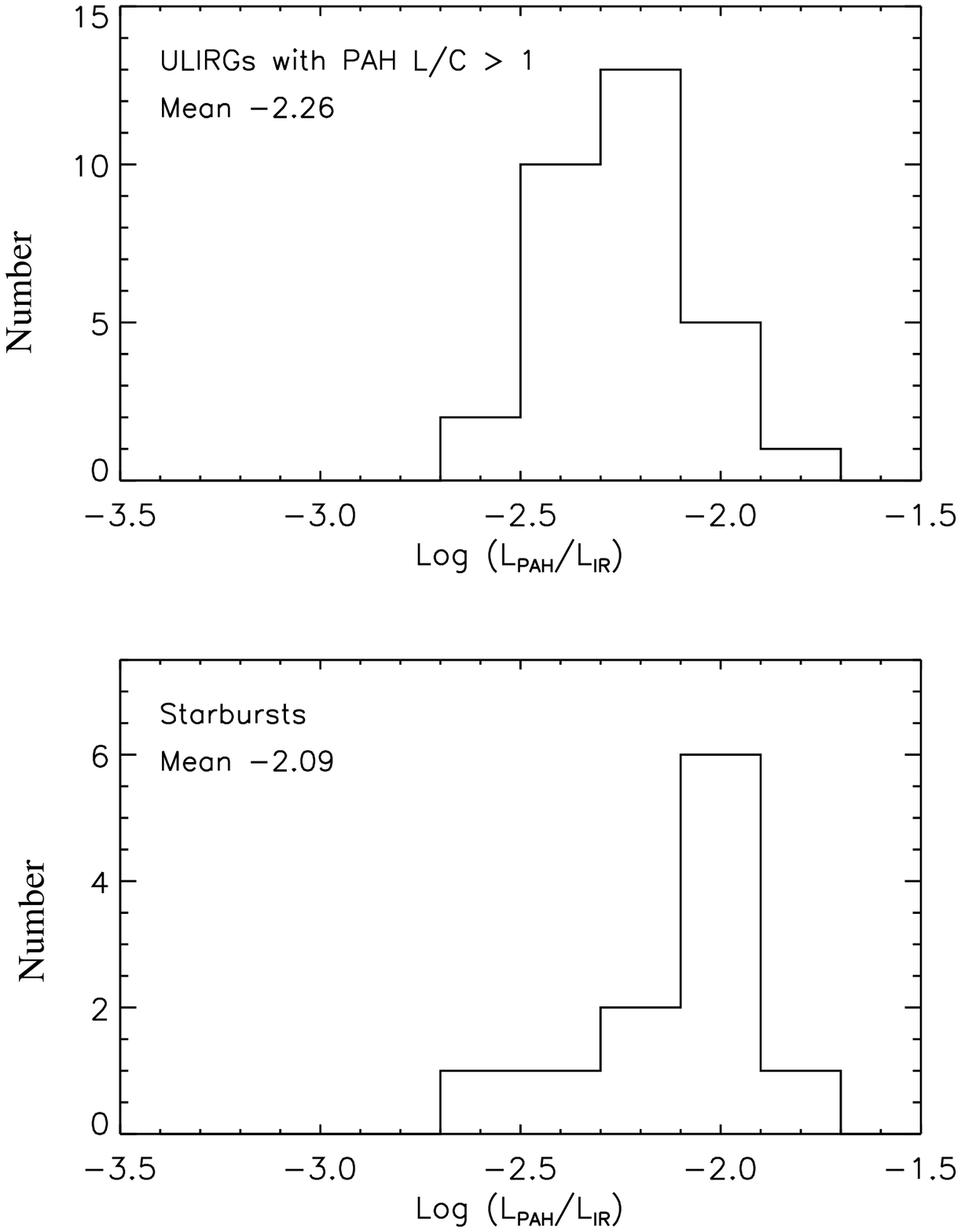]{Histograms showing the distribution of 
the L$_{\it PAH} /$ L$_{\it IR}$ ratio for ULIRG-starbursts and 
template starbursts.
\label{fig:rigopoulou5}}

\figcaption[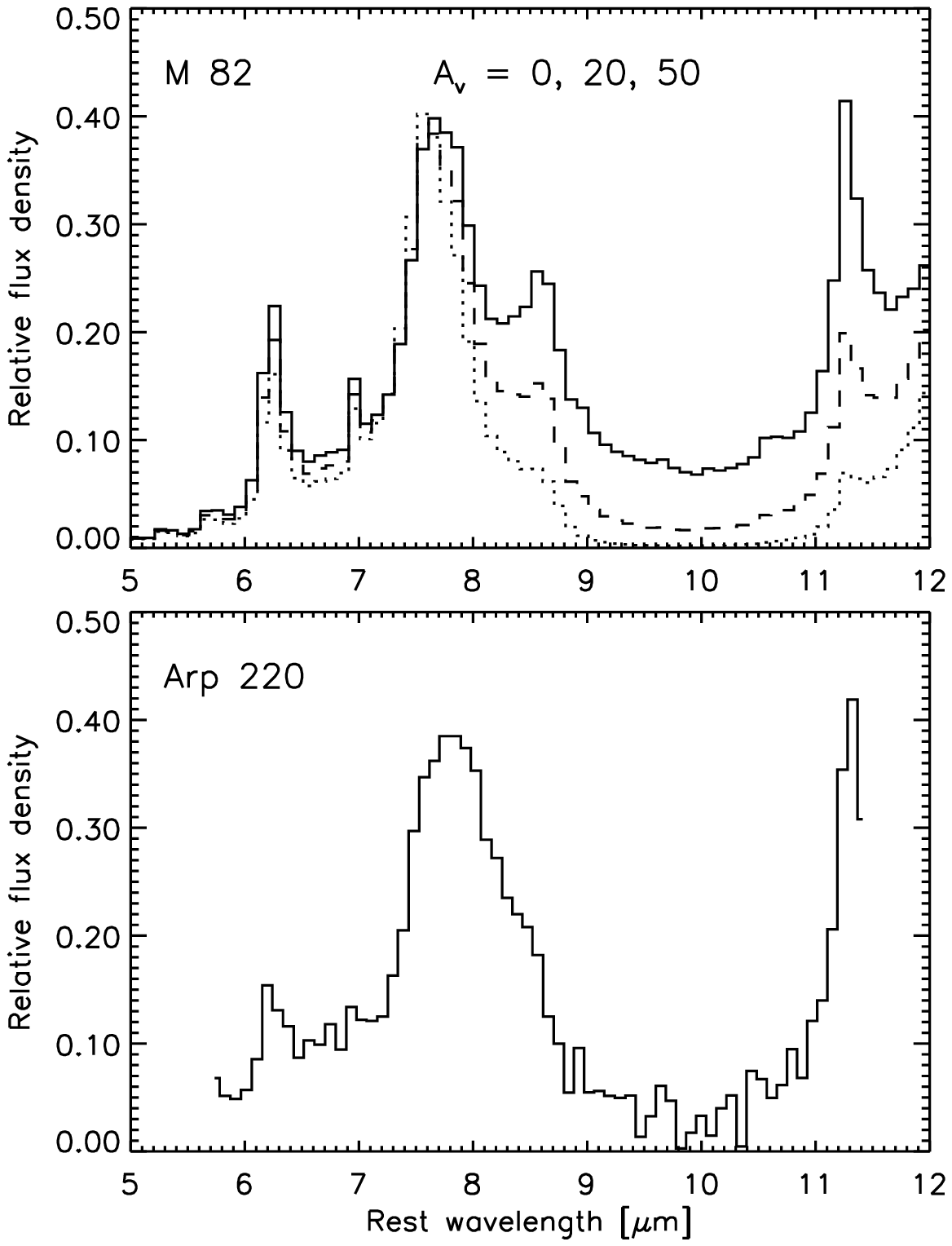]{Illustration of the effect of extinction on
PAH spectra. The top panel shows the spectrum of the starburst M 82, as
observed and assuming additional screen extinction of A$_V$=20 and 50.
The features at 6.2, 8.6, and 11.3$\mu$m are suppressed relative to the
one at 7.7$\mu$m to which the fluxes have been scaled for clarity.
The lower panel shows the spectrum of Arp 220, ressembling an obscured
starburst. The PHT-S long wavelength cutoff masks the fact that the
11.3$\mu$m feature of Arp220 is placed on a relatively stronger continuum.
\label{fig:rigopoulou6}}

\figcaption[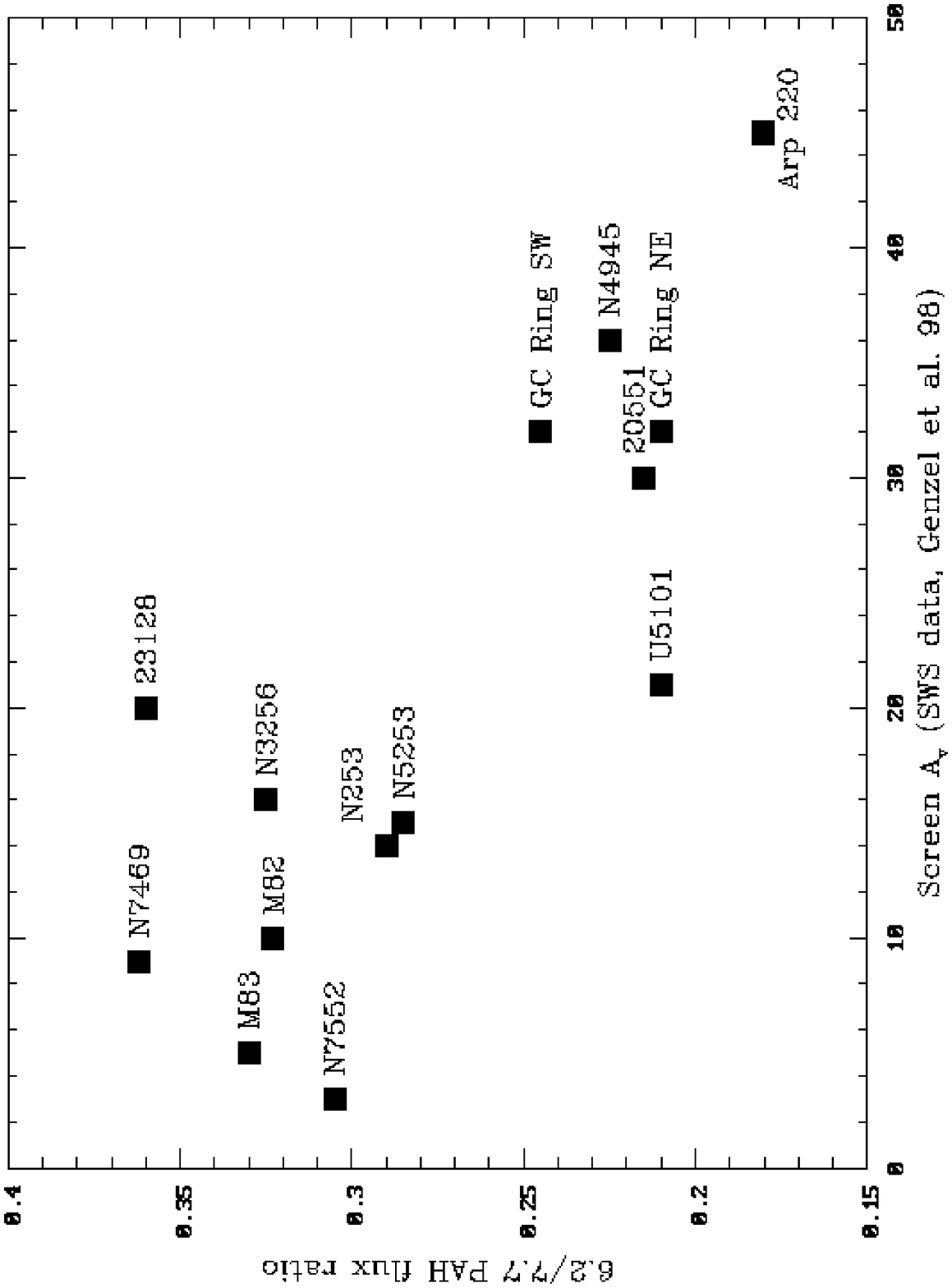]
{The 6.2$/$7.7 PAH ratio is plotted against screen extinction
A$_{\rm V}$ from independent SWS spectroscopy (Genzel et al. 1998)
\label{fig:rigopoulou7}}

\figcaption[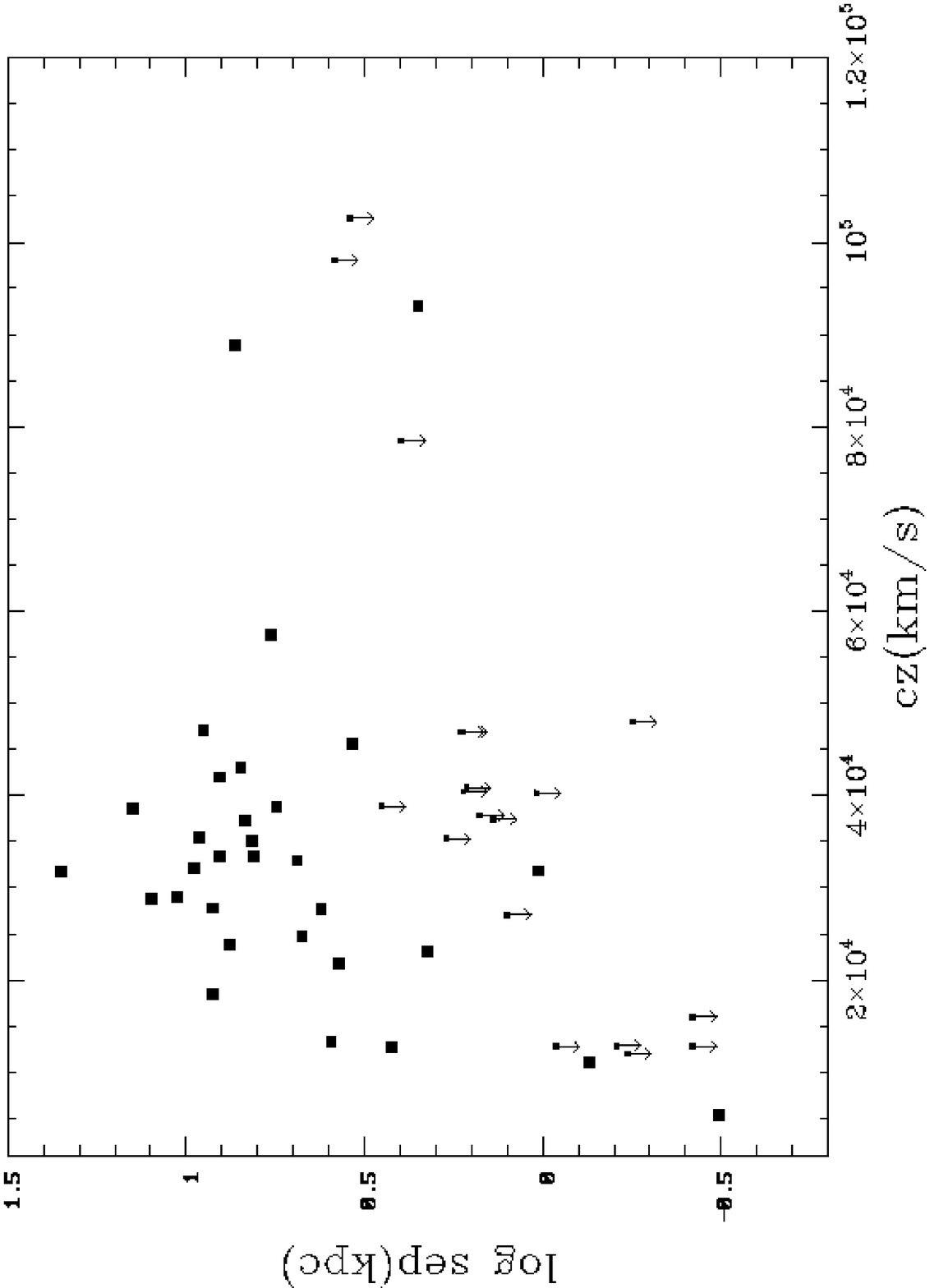]
{Projected nuclear separation as a function of redshift for the
ISO-ULIRGs. Open squares denote
ULIRGs with measured nuclear separations, limits denote ULIRGs with single
nuclei (or unresolved doubles) \label{fig:rigopoulou8}}

\figcaption[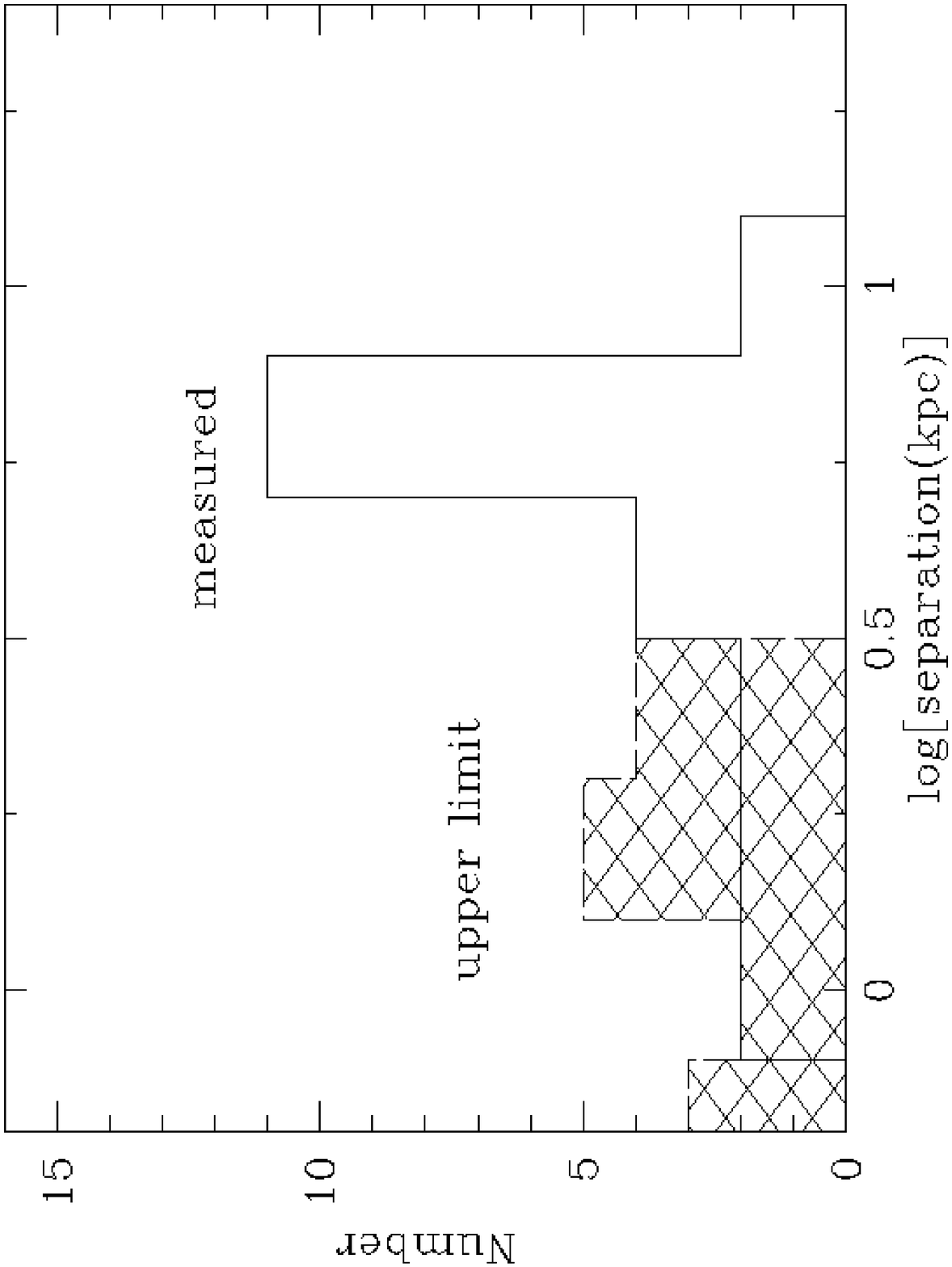]
{Histogram of the log of the projected linear separation (in kpc) for
the Southern ISO-ULIRGs.  The hatched area represents the galaxies with upper
limits in the their measured nuclear separations.
\label{fig:rigopoulou9}}

\figcaption[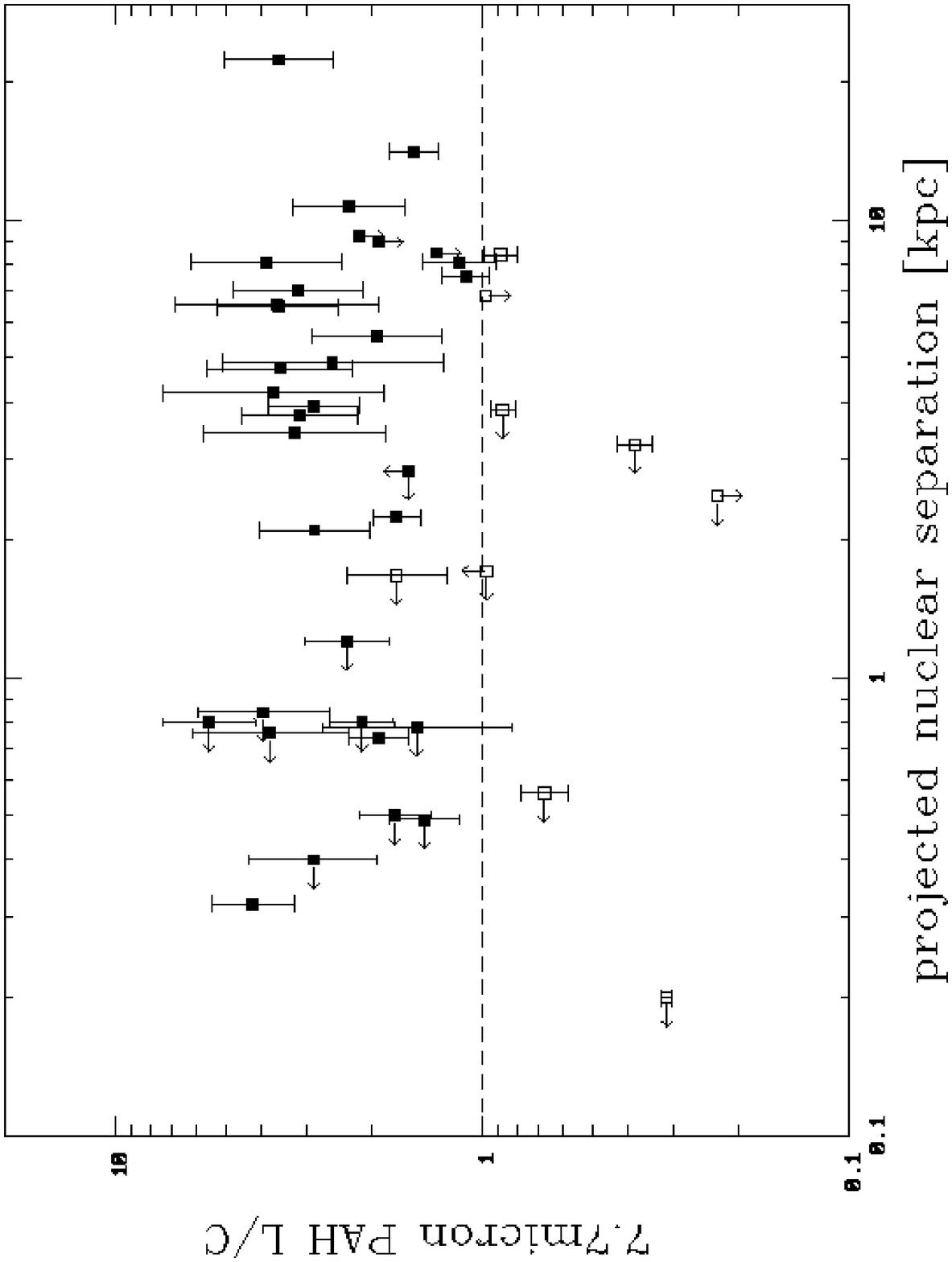]
{7.7 PAH L$/$C as a function of nuclear separation of the interacting
components of a ULIRG. Open symbols indicate AGN-ULIRGs, either according to our
classification or where there is evidence of an AGN from observations as other
wavelengths (although it
may not dominate the bolometric luminosity). Filled symbols indicate
starburst-ULIRGs. The dashed line indicated the adopted separation between
starbursts and AGN at L$/$C = 1 \label{fig:rigopoulou10}}

\figcaption[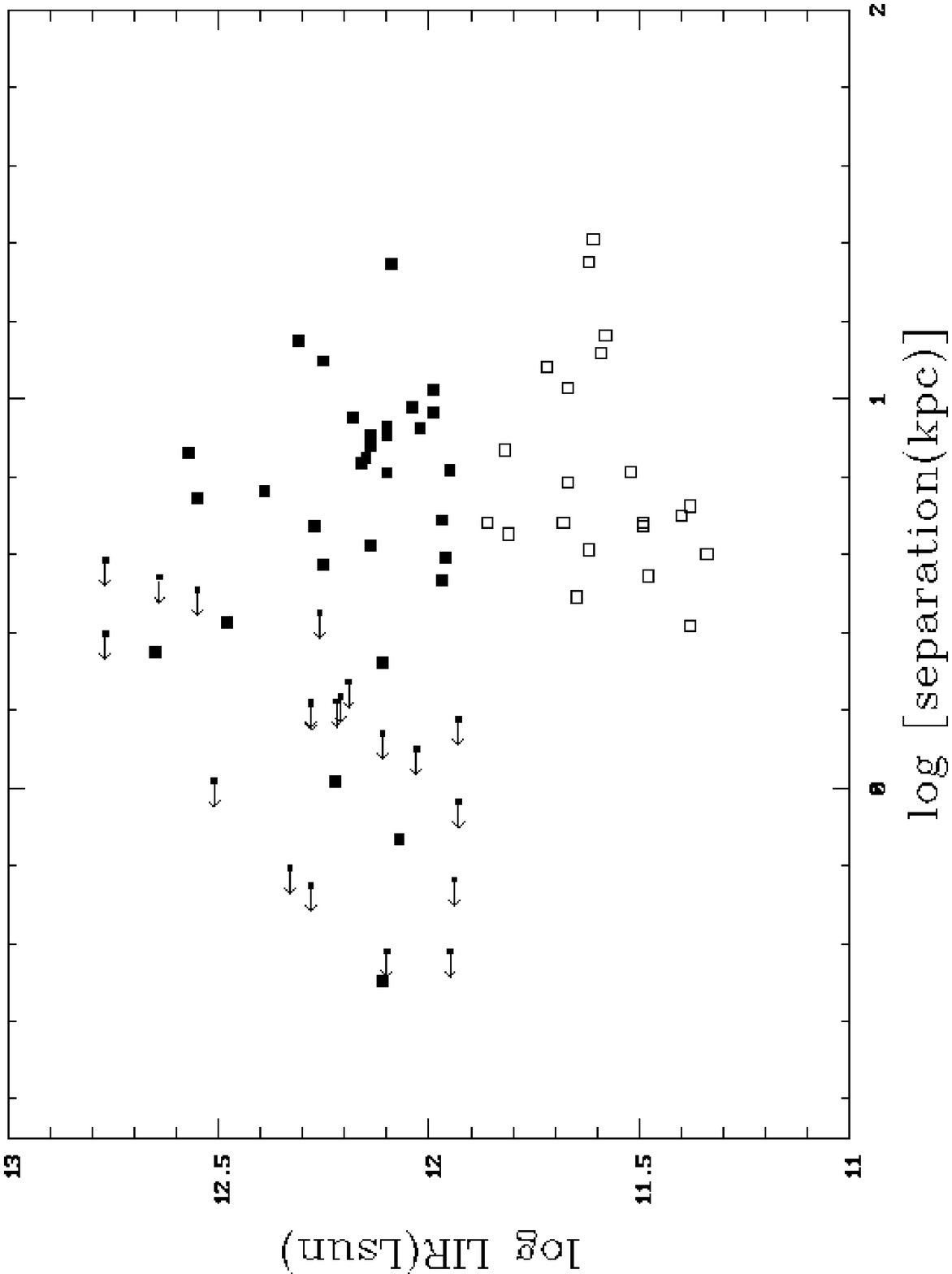]
{Infrared luminosity L$_{IR}$ as a function of projected nuclear
separation. Filled squares and upper limits correspond to our
ISO-ULIRG sample. Open squares are data for LIRGs taken from Gao et al. (1996)
and Gao and Solomon (1999) \label{fig:rigopoulou11}}

\figcaption[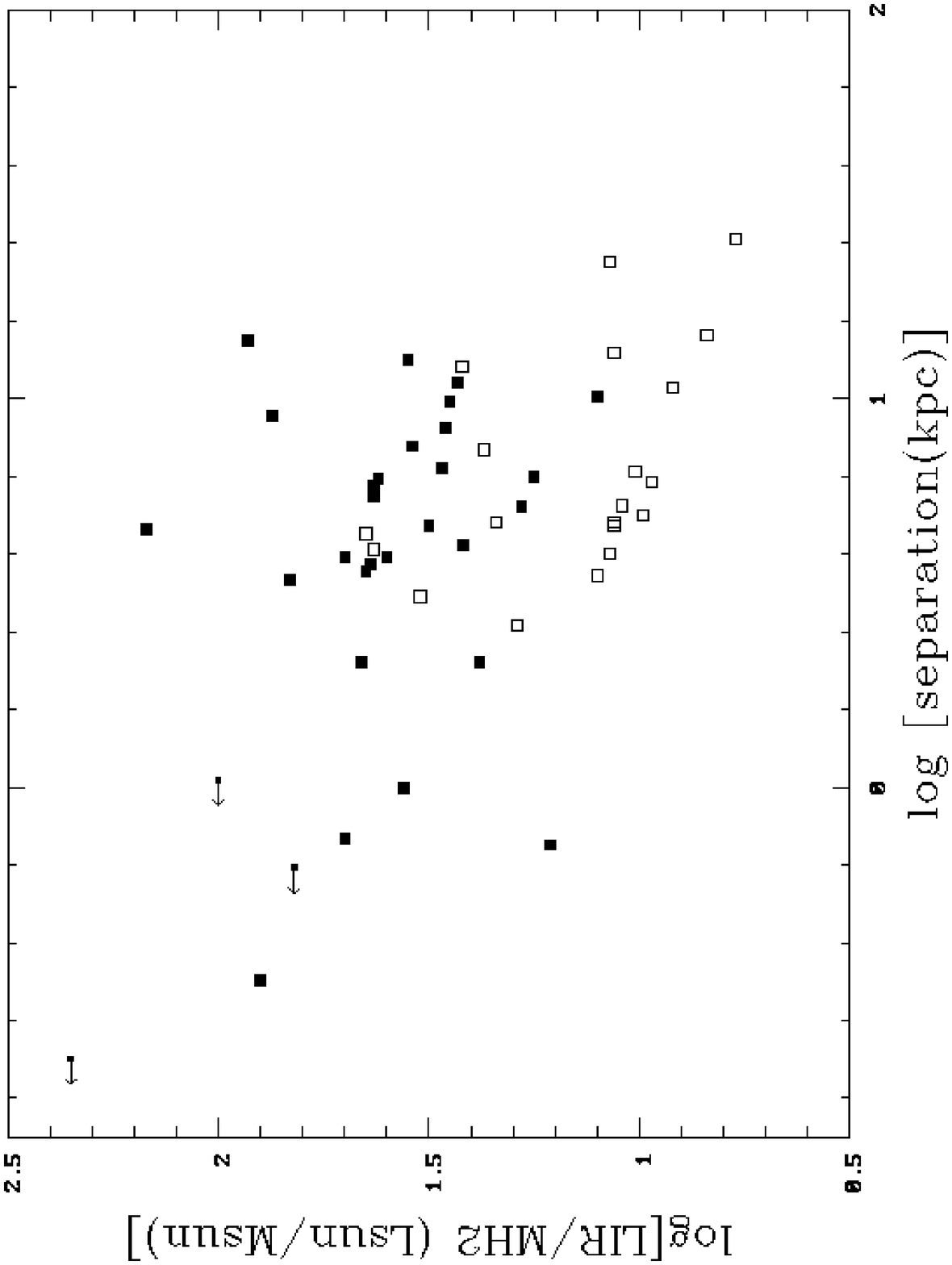]
{Star formation efficiency L$_{IR}/ $ M$_{H_{2}}$,
as a function of projected nuclear
separation.Filled squares and upper limits correspond to our
ISO-ULIRG sample. Open squares are data for LIRGs taken from Gao et al. (1996)
and Gao and Solomon (1999). Data for ULIRGs from Solomon et al. (1998) and Gao
and Solomon (1999) \label{fig:rigopoulou12}}

\figcaption[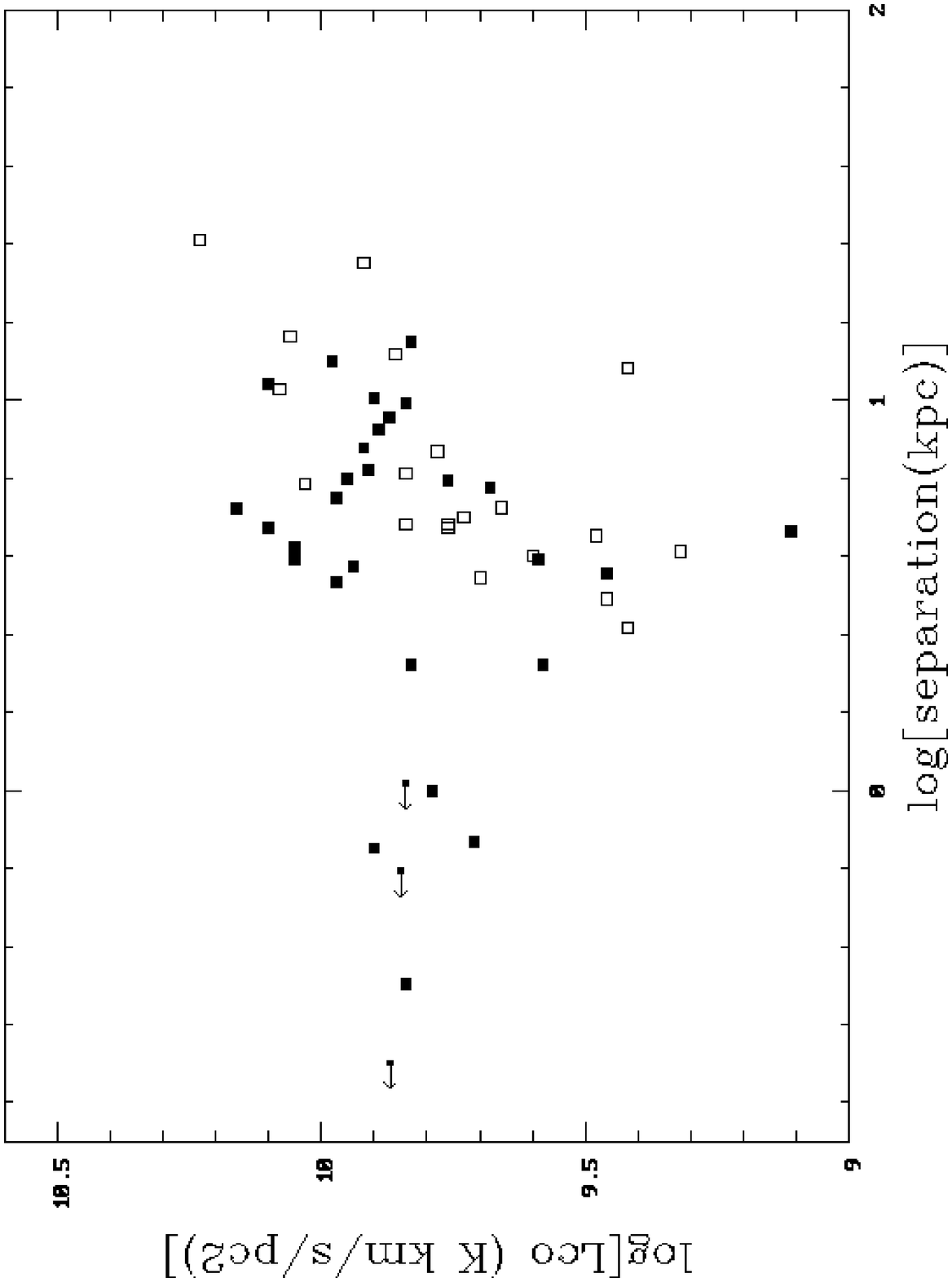]
{CO luminosity L$_{CO}$ as a function of projected nuclear
separation.Filled squares and upper limits correspond to our
ISO-ULIRG sample. Open squares are data for LIRGs taken from Gao et al. (1996)
and Gao and Solomon (1999). Data for ULIRGs from Solomon et al. (1998) and Gao
and Solomon (1999) \label{fig:rigopoulou13}}

\begin{deluxetable}{c c c c c c c}
\tablecolumns{7}
\tablewidth{33pc}
\tablenum{1}
\scriptsize
\tablecaption{ISO ULIRGs - PHOT-S Spectroscopy }
\tablehead{
\colhead{IRAS Name}&\colhead{cz}&\colhead{L$_{IR}$}&
\colhead{Cont$_{5.9}$}& \colhead{Cont$_{7.7}$}&
\colhead{F$_{7.7}$\tablenotemark{1a}}& \colhead{Line$/$Cont}\\
\colhead{} &\colhead{km$/$s} &\colhead{L$_{\odot}$}&
\colhead{x10$^{-2}$Jy}&
\colhead{x10$^{-2}$ Jy}& 
\colhead{x10$^{-2}$ Jy}&
\colhead{}
}

\startdata
23578-5307&37458&12.11&: 0.00\tablenotemark{1b}&0.00&$<$1.36&
NA\tablenotemark{1c}(0.0)\nl
00153+5454&33448&12.10&1.13&1.73&5.77& 3.340(0.821) \nl
0019-7426&28891&12.25&:0.00&: 0.00&$<$ 1.51&NA  (0.000) \nl
00262+4251&27800&12.02& 0.56& 1.46&5.93& 4.063  (1.444)\nl
00397-1312&78461&12.77& 6.44& 9.88&$<$ 2.26&  $<$ 0.229  (0.000) \nl
01003-2238&35286&12.19& 1.40&0.37 &3.10&   1.441  (0.385) \nl
01166-0844&35437&11.99&: 0.14&   0.84 &$<$ 1.81&   $<$ 2.161  (0.000) \nl
01199-2307&46887&12.21&: 0.07&: 0.74&2.08&$>$ 0.971  (0.000) \nl
01298-0744&40826&12.28&1.12&   2.07 &3.11&  1.503  (1.027) \nl
01355-1814&57410&12.39&: 0.028&$<$ 1.81&  :0.018&NA  (0.000) \nl
01388-4618&27068&12.03&1.12&   1.72 &6.77&   3.938  (0.713) \nl
01494-1845&47007&12.18&: -0.70&: -0.70  &3.60&  $>$ 1.951  (0.000) \nl
Mrk 1014&48900&12.50&2.52&3.87&$<$2.42& $<$ 0.626 (0.000)\nl
01569-2939&42030&12.14&: 0.14&    1.24 &1.42&    1.149  (0.398) \nl
02364-4751&29473&12.10&0.7&   1.45&5.06&    3.490  (0.736) \nl
02411+0354&43051&12.15&0.84&     1.49&4.73&    3.174  (0.703) \nl
03158+4227&40288&12.51&1.54&     2.36  &4.09&    1.729  (0.393) \nl
04063-3236&32949&11.97&0.46& 0.89& 2.28&   2.553  (1.203) \nl
04103-2838&35400&12.13& 1.54&2.36&3.26&  1.378  (0.337) \nl
04114-5117&37345&12.16& 0.28&0.93&$<$ 9.06& $<$ 0.975  (0.000) \nl
06009-7716&35070&11.95&0.31&0.95&3.44&  3.632  (1.105) \nl
06035-7102&23823&12.14&4.89&7.51&8.32&  1.108  (0.258) \nl
06206-6315&27713&12.14&  1.23&1.44&5.31&  3.687  (2.206) \nl
06301-7934&46891&12.28&: 0.14&: 0.34&$<$ 1.06&NA  (0.000) \nl
06361-6217&47967&12.28&  1.40&2.15&1.45&   0.676  (0.258) \nl
UGC 5101&12000&11.94& 5.90& 8.78&18.6&  2.123  (0.346)\nl
09463+8141\tablenotemark{1d}&46378&12.20& & & & \\
12112+0305&21788&12.25&  1.68&2.58&8.08&  3.135  (0.634)\nl
Mrk 231&12660&12.48&60.6&    91.9&28.9&  0.314  (0.021)\nl
Mrk 273&11132&12.07&7.62& 10.1 &19.3& 1.908  (0.322)\nl
14348-1447&24732&12.27&1.26&1.93&6.86&   3.550  (0.797)\nl
15250+3609&16000&11.95&2.80&4.45&12.8&   2.880 (0.695)\nl
Arp 220&5450&12.11& 7.00& 9.84&41.4&    4.205  (0.447)\nl
16334+4630&57265&12.34&:0.28&: 0.28&$<$ 1.21&NA  (0.000) \nl
16474+3430&33418&12.10&0.98&1.58&5.66&  3.586  (0.657) \nl
16487+5447&31293&12.11& 0.28&0.88&1.93&  2.196  (1.111) \nl
NGC 6240&7339&11.78&  10.4&16.2&42.0&  2.586  (0.238) \nl
17028+5817&31779&12.09& 0.32&1.01&3.60&  3.580  (1.087) \nl
17068+4027&53700&12.27& 0.69&1.24&1.75&  1.409  (0.583) \nl
17179+5444&44211&12.16& 0.70&1.20&$<$ 1.66&  $<$ 1.386  (0.000)\nl
17208-0014&12900&12.33&  3.22&4.94&27.4&    5.549  (1.006) \nl
18443+743&40395&12.22&0.98&1.48& 2.53&    1.708  (0.550) \nl
18470+3233&23626&12.01&: 0.00&: 0.00& $<$ 0.651&NA  (0.000) \nl
18531-4616&42202&12.17&0.42&1.02&2.65&   2.597  (1.283) \nl
19254-7245&18500&12.02&7.00&9.50&8.46&    0.891  (0.175) \nl
19420+4556&35356&11.99&0.28&67.9&2.51&  3.698  (2.024) \nl
19458+0944&29980&12.28& 0.70&1.55&3.79&   2.447  (1.150)\nl
20049-7210&37893&11.93&: 0.00&0.35&1.32&  3.771  (2.448) \nl
20100-4156&38848&12.55& 1.40&2.15&4.14&   1.926  (0.707) \nl
20446-6218&32078&12.04&: 0.00&: 0.5&$<$ 2.11&NA (0.000) \nl
20551-425&12788&11.98& 4.53&6.96&16.3&    2.339  (0.458) \nl
21396+3623&42000\tablenotemark{1e}&11.99&0.64&1.31 &3.04&   2.311  (0.615) \nl
22055+3024&38041&12.15&: 0.28&: 0.68&$<$ 1.97&NA (0.000) \nl
22491-1808&23170&12.11&0.84&1.54&4.39&   2.854  (1.603)\nl
23060+0505&52200&12.41&9.10&14.0&$<$ 1.21&   $<$ 0.087  (0.000) \nl
23129+2548&53700&12.38&0.42&0.52&1.85&    3.552  (1.928) \nl
23128-5919&13371&11.96&4.17&6.40&18.4&   2.870  (0.504) \nl
23230-6926&31870&12.22&: 0.42&: 0.42&$<$ 3.78&NA (0.000) \nl
23253-5415&38838&12.26& 0.31&:0.35&1.66&$>$ 1.588  (0.000) \nl
23327+2913&32000 &12.04 &: 0.14&: 0.59&$<$2.27&NA(0.0)\nl
23365+3604&19340 &12.09 &1.26&1.93&8.66&4.479(1.055)\nl
23389-6139&27806&12.10&: 0.42&: 0.42&2.89 &$>$ 1.334(0.000) \nl
\enddata

\tablenotetext{1a}{7.7 PAH mean peak flux}
\tablenotetext{1b}{: the derived 5.9 $\mu$m or 7.7 $\mu$m continuum values
 are within 3 $\sigma$ from zero}
\tablenotetext{1c}{NA: the 7.7 $\mu$m L$/$C cannot be defined}
\tablenotetext{1d}{The 7.7 $\mu$m PAH feature is present. However, 
no PAH continuum estimates since the off position contains
a star which contaminates all background measurements} 
\tablenotetext{1e}{velocity derived from PAH features (inconsistent with
cz value of 29304 km$/$s quoted by Strauss et al. (1992)}
\tablecomments{
Column (1): source name, 
column (2): velocity cz,
column (3): L$_{IR}$ luminosity (see text), 
column (4): continuum at 5.9 $\mu$m,
column (6): continuum at 7.7 $\mu$m,
column (7): 7.7$\mu$m PAH peak flux,
column (8): line-to-continuum ratio of
the 7.7 $\mu$m PAH emission feature and error (in parentheses).}
\end{deluxetable}

\begin{deluxetable}{c c c c c c c}
\tablecolumns{7}
\tablewidth{33pc}
\tablenum{2}
\scriptsize
\tablecaption{ISO AGN - PHOT-S Spectroscopy }
\tablehead{
\colhead{Name}&\colhead{cz}&\colhead{L$_{IR}$}&
\colhead{Cont$_{5.9}$}& \colhead{Cont$_{7.7}$}&
\colhead{F$_{7.7}$\tablenotemark{2a}}& \colhead{Line$/$Cont}\\
\colhead{} &\colhead{km$/$s} &\colhead{L$_{\odot}$}&
\colhead{x10$^{-2}$Jy}&
\colhead{x10$^{-2}$ Jy}& 
\colhead{x10$^{-2}$ Jy}&
\colhead{}
}

\startdata
IZw1&18342&11.87&17.3&22.1&$<$4.85&$<$0.219(0.0)\nl
Mrk 1&4780&10.53&2.09&8.83&$<$2.71&$<$0.307(0.0)\nl
NGC 1068&1148 &11.29&866&145&233&0.161(0.018)\nl
4U0241+62&13200&11.44&24.5&30.7&$<$2.86&$<$0.093(0.0)\nl
NGC 1275&5264 &11.21&11.5&21.4&$<$3.76&$<$0.176(0.0)\nl
NGC 1365&1636 &11.07 &55.9&96.6&221&2.290(0.129)\nl
PG0804+761&30000&11.52&7.14&7.02&$<$1.21&$<$0.172(0.0)\nl
NGC 3783&2550 &10.39 &20.7&28.3&$<$3.61&$<$0.128(0.0)\nl
NGC 4151&995&10.20&88.2&109&$<$8.45&$<$0.077(0.0)\nl
3C273&47500 &12.69&18.5&19.6&$<$2.43&$<$0.124(0.0)\nl
M 87&1282 &8.84 &10.2&7.03&$<$1.51&$<$0.214(0.0)\nl
CenA&547 &9.98 &100&122&153&1.258(0.066)\nl
Mrk 463&14904 &11.71 &23.8&29.8&6.27&0.211(0.034)\nl
Circinus&436 &9.87 &450&623&1002&1.630(0.062)\nl
NGC 5506&1853 &10.36 &64.6&71.3&31.2&0.437(0.032)\nl
NGC 5643&1199 &10.25 &7.1&17.8&11.5&0.645(0.122)\nl
PG1613+658&38700 &11.91 &3.92&6.02&$<$1.06&$<$0.176(0.0)\nl
PG1700+518&87600 &12.59 &3.5&3.75&$<$1.97&$<$0.524(0.0)\nl
PKS2048-57&3402 &10.79 &23.8&43.3&$<$6.07&$<$0.140(0.0)\nl
HB21219-1757&33900 &12.04 &6.46&9.91&$<$1.51&$<$0.152(0.0)\nl
PG2130+099&18630 &11.36 &8.02&9.41&$<$1.81&$<$0.193(0.0)\nl
NGC 7469&4892 &11.56 &27.4&53.9&53.8&0.997(0.061)\nl
NGC 7582&1575 &10.80 &33.2&50.9&127&2.484(0.158)\nl
\enddata

\tablenotetext{2a}{7.7 PAH mean peak flux}
\tablecomments{
Column (1): source name, 
column (2): velocity cz,
column (3): L$_{IR}$ luminosity (see text), 
column (4): continuum at 5.9 $\mu$m,
column (6): continuum at 7.7 $\mu$m,
column (7): 7.7$\mu$m PAH peak flux, 
column (8): line-to-continuum ratio of
the 7.7 $\mu$m PAH emission feature and error (in parentheses).}

\end{deluxetable}

\begin{deluxetable}{c c c c c c c}
\tablecolumns{7}
\tablewidth{33pc}
\tablenum{3}
\scriptsize
\tablecaption{ISO starbursts - PHOT-S Spectroscopy }
\tablehead{
\colhead{IRAS Name}&\colhead{cz}&\colhead{L$_{IR}$}&
\colhead{Cont$_{5.9}$}& \colhead{Cont$_{ 7.7}$}&
\colhead{F$_{7.7}$\tablenotemark{3a}}& \colhead{Line$/$Cont}\\
\colhead{} &\colhead{km$/$s} &\colhead{L$_{\odot}$}&
\colhead{x10$^{-1}$Jy}&
\colhead{x10$^{-1}$ Jy}&
\colhead{x10$^{-1}$ Jy}&
\colhead{}
}

\startdata
NGC 253&245 &10.44&15.6&33.1&100&3.018(0.192)\nl
IC 342&34 &9.56 &4.35&9.80&27.7&2.826(0.190)\nl
M 82\tablenotemark{3b}&203 &10.56 &35.8&67.8&284&4.195(0.110)\nl
NGC 3256&2738&11.55&4.45&7.89&33.0&4.183(0.243)\nl
NGC 4945&560 &10.20 &8.33&7.34&78.4&10.69(0.827)\nl
M 83&516 &9.81 &6.24&13.9&44.4&3.194(0.245)\nl
NGC 5253&404&9.29&3.29&5.44&2.71&0.498(0.047)\nl
NGC 6764&2416 &10.36 &0.406&1.06&1.38&1.303(0.180)\nl
NGC 6946&48&9.83 &2.83&4.25&22.7&5.350(0.357)\nl
NGC 7552&1585 &10.99&3.77&7.68&22.2&2.885(0.189)\nl
NGC 891\tablenotemark{3c}&528&9.84 &1.61&1.80&7.26&4.023(0.277)\nl
M 81&-34 &8.58 &3.75&2.91&$<$0.34&$<$0.115(0.00)\nl
M 51\tablenotemark{3c} &463 &9.75 &1.01&1.70&2.39&1.411(0.149)\nl
NGC 4569\tablenotemark{3c}&-235 &9.90 &0.82&1.46&1.86&1.276(0.185)\nl
\enddata

\tablenotetext{3a}{7.7 PAH mean peak flux}
\tablenotetext{3b}{SWS AOT1 spectrum}
\tablenotetext{3c}{Normal Galaxies}
\tablecomments{
Column (1): source name, 
column (2): velocity cz,
column (3): L$_{IR}$ luminosity (see text), 
column (4): continuum at 5.9 $\mu$m,
column (6): continuum at 7.7 $\mu$m,
column (7): 7.7$\mu$m PAH peak flux,
column (8): line-to-continuum ratio of
the 7.7 $\mu$m PAH emission feature and error (in parentheses).}

\end{deluxetable}

\begin{deluxetable}{c c c c c}
\tablecolumns{3}
\tablewidth{33pc}
\tablenum{4}
\scriptsize
\tablecaption{Comparison of the Averaged Spectra}
\tablehead{
\colhead{}&
\colhead{Cont$_{5.9}$\tablenotemark{4a}}& \colhead{Cont$_{7.7}$}&
\colhead{F$_{7.7}$}& \colhead{Line$/$Cont}\\
\colhead{}& \colhead{x10$^{-2}$Jy}&
\colhead{x10$^{-2}$ Jy}& \colhead{x10$^{-2}$ Jy}&
\colhead{}
}
\startdata
ULIRGs&0.4&0.61&1.27&2.061(0.103)\nl
starbursts&0.5&0.93&2.77&2.983(0.161)\nl
AGN&7.82&8.76&0.4&0.212(0.003)\nl
\enddata
\tablenotetext{4a}{5.9 $\mu$m continua relative to the 60 $\mu$m flux}
\tablecomments{For the {\it averaged} AGN spectrum the following galaxies 
were used: IZw1, Mrk 1, NGC 1068, NGC 1275, NGC 3783, NGC 4151, 3C273, MRK 463,
NGC 5506, NGC 5643, PG1613$+$658, PG1700$+$518, PKS2048--57, HB21219-1757,
PG2130$+$099\\ 
The following galaxies contribute to the {\it averaged} starburst spectrum:
NGC 253, IC342, M82, NGC 3256, NGC 4945, M83, NGC 5253, NGC 6946, NGC 7552
} 
\end{deluxetable}

\begin{deluxetable}{c c c c c c c c}
\tablecolumns{8}
\tablewidth{33pc}
\tablenum{5}
\scriptsize
\tablecaption{ISO ULIRGs - Infrared imaging }
\tablehead{
\colhead{IRAS Name}&\colhead{m$_{K}$}&
\colhead{m$_{J}$}&\colhead{Ap(")}&
\colhead{Sep(")}&\colhead{Sep(kpc)}&
\colhead{Clas}&\colhead{Notes}}

\startdata
23578--5307&13.24&14.92&10&$<$0.7&$<$1.4&merger completed&1\nl
F00183--7111&14.43&16.27&10&$<$1.0&$<$3.8&merger completed&1\nl
00188--0856N\tablenotemark{5a}&12.96&15.05&5& & & &2\nl
00188--0856S&14.73&15.92&5& & & &2\nl
00188--0856 &12.65&14.49&16&7.0&14.0&interacting&1\nl
0019--7426N&12.38&14.03&5&&& &2\nl
0019--7426S&15.05&16.82&5&&& &2\nl
0019--7426 & 12.13&13.76&16&7.9&12.5&interacting&1\nl
00397--1312&13.99&16.04&10&$<$0.7&$<$2.5&merger completed&1\nl
00406--3127&15.00&16.25&10&$<$0.9&$<$3.5&merger completed&1\nl
01003--2238&14.10&15.83&10&$<$0.5&$<$1.9&relaxed&1\nl
01166--0844W&14.68&16.35&3&& & &2\nl
01166--0844E&14.93&16.52&3&& & &2\nl
01166--0844 &13.7&15.16&12&4.9&9.2&interacting&1\nl
01199--2307&14.69&16.07&10&$<$0.7&$<$1.7&&1\nl
01298--0744&14.28&15.99&10&$<$0.8&$<$1.6&merger completed&1\nl
01355--1814\tablenotemark{5b}&14.22& &10&2.1&5.8&interacting&1\nl
01388--4618&12.23&13.32&10&$<$0.8&$<$1.3&&1\nl
01494--1845N&13.87&15.86&2&&& &2\nl
01494--1845S&16.27&18.15&2&&& &2\nl
01494--1845 &13.42&15.18&10&3.8 &8.9&interacting &1\nl
02411+0354 W&14.49 &15.97&1.5&&& &1\nl
02411+0354 Main&14.24&15.86&1.5&&& &2\nl
02411+0354 S&14.97&16.46&1.5& && &2\nl
02411+0354 &12.67&14.37&12&3.2 &7.0&interacting &1\nl
F02455--2220&14.63&16.34&10&2.0&7.3&merger completed&1\nl
03521+0028 W&14.76&16.74&1.0& && &2\nl
03521+0028 E&15.29&16.92&1.0&&& &2\nl
03521+0028 &13.63&15.32&10&1.5&3.4&interacting&1\nl
04063--3236 SW&14.93&16.22&1.5&& & &2\nl
04063--3236 NE&14.25&16.02&1.5&& & &2\nl 
04063--3236 &13.45&14.83&10&2.7 &4.9&interacting &1\nl
04114--5117 E&14.19&15.90&2& && &2\nl
04114--5117 SW&15.12&16.39&2& && &2\nl
04114--5117 &13.58&14.70&10&3.5 &6.8&interacting &1\nl
05189--2524&10.14&12.62&10&$<$0.5&$<$0.38&relaxed&1\nl
06009--7716W&14.19&15.40&2&& &&2\nl
06009--7716E&14.78&15.69&2&&& &2\nl
06009--7716 &13.36&14.45&10&3.5 &6.5&interacting&1\nl
06206--6315SW&13.64& &1.5&&& &2\nl
06206--6315NE&14.46& &1.5&&& &2\nl
06206--6315 &12.60& &10&2.7 &4.2&interacting&1\nl
06301--7934&13.64&15.39&15&$<$0.7&$<$1.7&&1\nl
20049--7210&14.08&15.01&10&$<$0.8&$<$1.5&&1\nl
20446--6218N&13.21&14.61&3& & &&2\nl
20446--6218S&14.96&15.86&3& & &&2\nl
20446--6218 &12.54&13.71&15&5.5&9.5&interacting &1\nl
23253--5415&12.75&13.58&12&$<$1.4 &$<$2.8&merger completed&1\nl
23389--6139S&13.68&15.28&3& & &&2\nl
23389--6139N&16.82&16.12&3& & &&2\nl
23389--6139 &13.13&14.45&10&5.5 &8.5&interacting&1\nl
23529--2119&14.48&16.68&10&$<$0.7&$<$3.2&merger completed&1\nl  
\enddata
\tablenotetext{5a}{the double nuclei are distinguished by their orientation on
the images, north, south etc.}
\tablenotetext{5b}{K$_{\rm s}$ integration time : 8 mins}
\tablecomments{ 1: For single nuclei objects magnitudes are quoted for a 10"
aperture.\\
2: For double nuclei objects magnitudes are quoted for the two
nuclei separately: usually the magnitude for each nucleus 
is determined through a 5" or 3" apertures.
The total magnitude quoted for double-nuclei systems corresponds to apertures
of 10" or 12" or 16" (depending on the sizes of the objects).}
\tablecomments{ Column (1): source name, columns (2)-(3):
K$_{\rm s}$ and J magnitudes, column (4): the aperture used to derive the
magnitudes, column (5)-(6): angular('') and linear (kpc) separations, 
column(7):ULIRG interaction stage classification (see section 5.2),  
column (8): notes for the derived magnitudes. 
}
\end{deluxetable}

\begin{deluxetable}{c c c c c}
\tablecolumns{5}
\tablewidth{0pc}
\tablenum{6}
\scriptsize
\tablecaption{ISO ULIRGs - additional separations}
\tablehead{
\colhead{IRAS Name}& \colhead{Sep(")}& 
\colhead{Sep(kpc)}&\colhead{Ref}}
\startdata
00153+5454&4.5&8.08&Murphy\nl
01569-2939&3.75&8.09&Duc\nl
03158+4227&$<$0.5&$<$1.0&Murphy\nl
03538-6432&0.6&2.2&NICMOS\nl
06035-7102&5.6&7.5&NICMOS\nl
06361-6217&$<$0.2&$<$0.6&NICMOS\nl
UGC 5101&$<$0.8&0.6&Carico\nl
12112+0305&3.0&3.7&Carico\nl
Mrk 231&$<$0.2&$<$0.2&Lai\nl
Mrk 273&1.1&0.7&Majewski\nl
14348-1447&3.4&4.7&Carico\nl
15250+3609&$<$0.4&$<$0.4&Majewski\nl
Arp 220&0.9&0.3&Graham\nl
16334+4630&4.4&12.0&Gao\nl
16474+3430&3.6&6.5&Murphy\nl
16487+5447&3.1&5.3&Murphy\nl
17028+5817&13.0&22.0&Murphy\nl
17208-0014&$<$0.8&$<$0.6&Murphy\nl
18443+7433&$<$0.8&$<$1.7&Murphy\nl   
19254-7245&7.8&8.4&Duc\nl
20100-4156&2.7&5.6&NICMOS\nl
20551-4250&$<$1.2&$<$0.9&NICMOS\nl
21396+3623&6.7&10.0&Murphy\nl
22491-1808&1.6&2.1&Carico\nl
23128-5919&4.9&3.9&NICMOS\nl
23230-6926&0.6&1.04&NICMOS\nl
\enddata
\tablecomments{column (1): source name, column (2)-(3): angular('') and linear
(kpc) separation, column (4): references, Carico: Carico et al. 1990, Duc: Duc,
Mirabel, Maza 1997,  Gao: Gao and Solomon 1999,  Graham: Graham et al. 1990,
Lai: Lai et al. 1998, Majewski: Majewski et al. 1993, Murphy: Murphy et al.
(1996), NICMOS: NICMOS archive (Borne et al.)  
}
\end{deluxetable}

\begin{deluxetable}{c c c c c}
\tablecolumns{5}
\tablewidth{0pc}
\tablenum{7}
\scriptsize
\tablecaption{ISO ULIRGs - Interaction Stages}
\tablehead{
\colhead{ }&
\colhead{Fully Relaxed}& \colhead{Merger Completed}&
\colhead{Interacting Pairs}}
\startdata
Number of galaxies&2&6-7&12\nl
Percentage&7.4\%&22\%&47\%\nl
\enddata
\end{deluxetable}

\begin{figure}[hbtp]
\figurenum{1}
\psfig{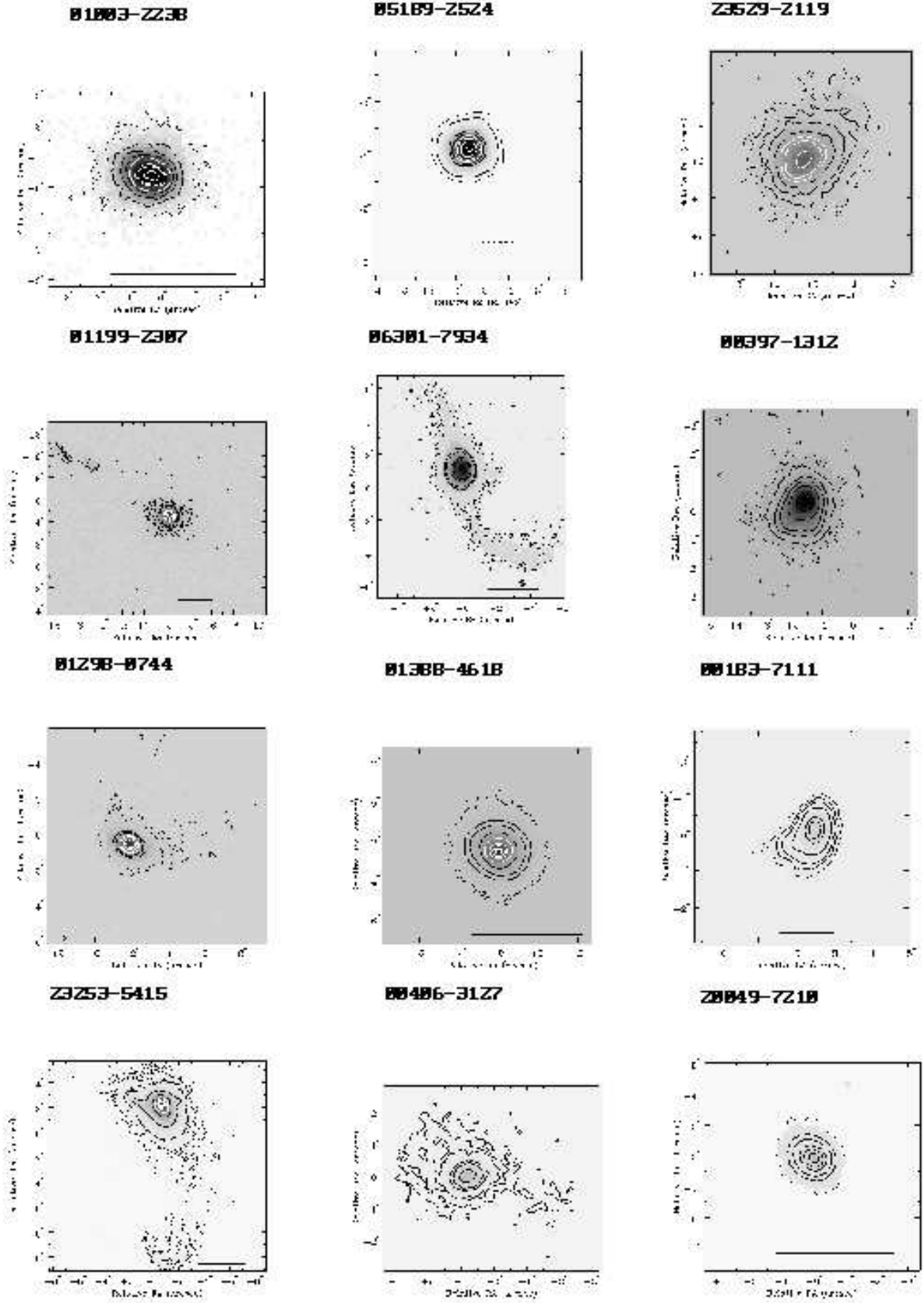}
\caption{K contours}
\end{figure}

\begin{figure}[hbtp]
\setcounter{figure}{1}
\figurenum{1}
\psfig{file=rigopoulou1b.eps,height=22cm,width=18cm}
\caption{K contours, continued}
\end{figure}

\begin{figure}[hbtp]
\setcounter{figure}{1}
\figurenum{1}
\psfig{file=rigopoulou1c.eps,width=18cm}
\caption{K contours, continued}
\end{figure}

\begin{figure}[hbtp]
\figurenum{2}
\psfig{file=rigopoulou2a.eps}
\caption{ISOPHOT-S Spectroscopy}
\end{figure}

\begin{figure}[hbtp]
\figurenum{2}
\setcounter{figure}{2}
\psfig{file=rigopoulou2b.eps}
\caption{ISOPHOT-S Spectroscopy, continued}
\end{figure}

\begin{figure}[hbtp]
\figurenum{3}
\setcounter{figure}{3}
\psfig{file=rigopoulou3.eps}
\caption{ISOPHOT-S Spectroscopy, AGN}
\end{figure}

\begin{figure}[hbtp]
\figurenum{4}
\setcounter{figure}{4}
\psfig{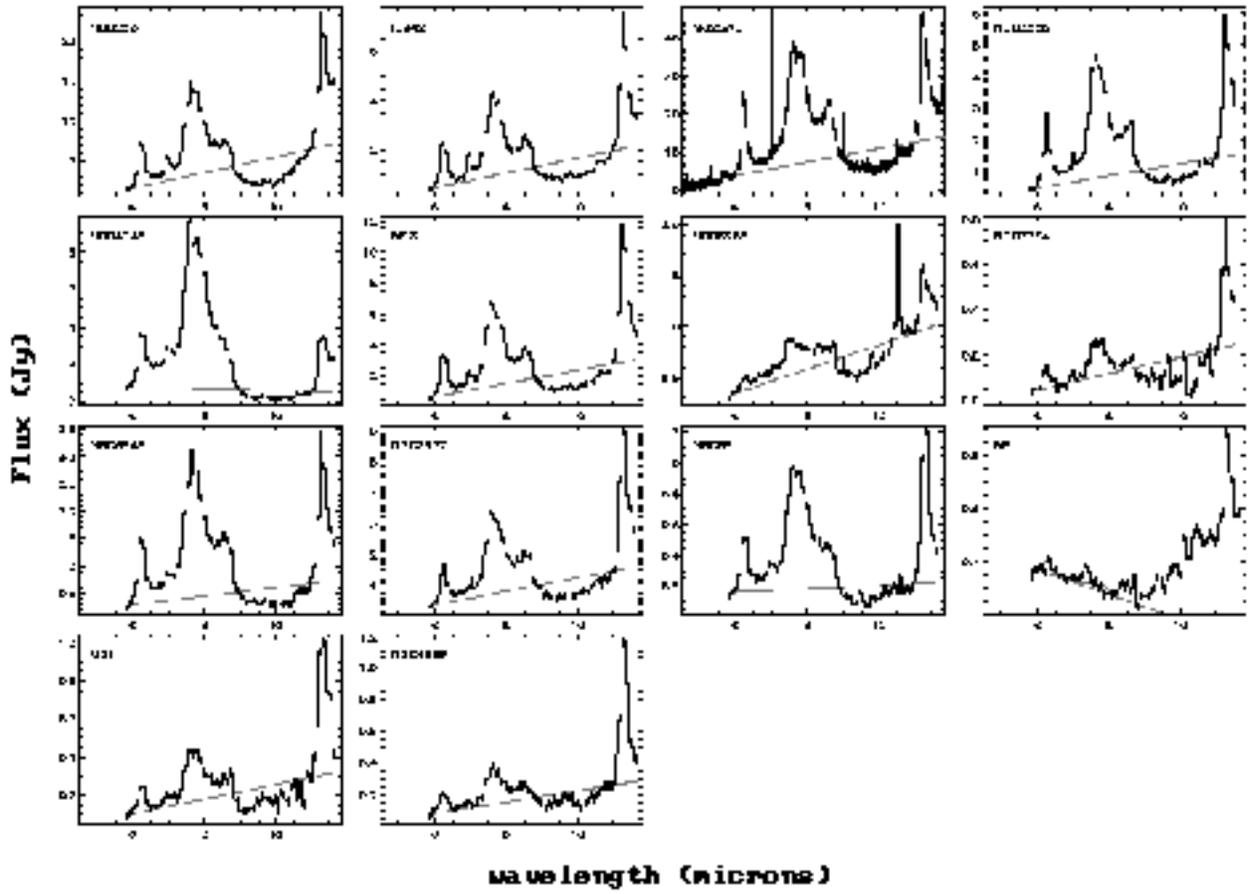}
\caption{ISOPHOT-S Spectroscopy, starbursts}
\end{figure}

\begin{figure}[hbtp]
\figurenum{5}
\psfig{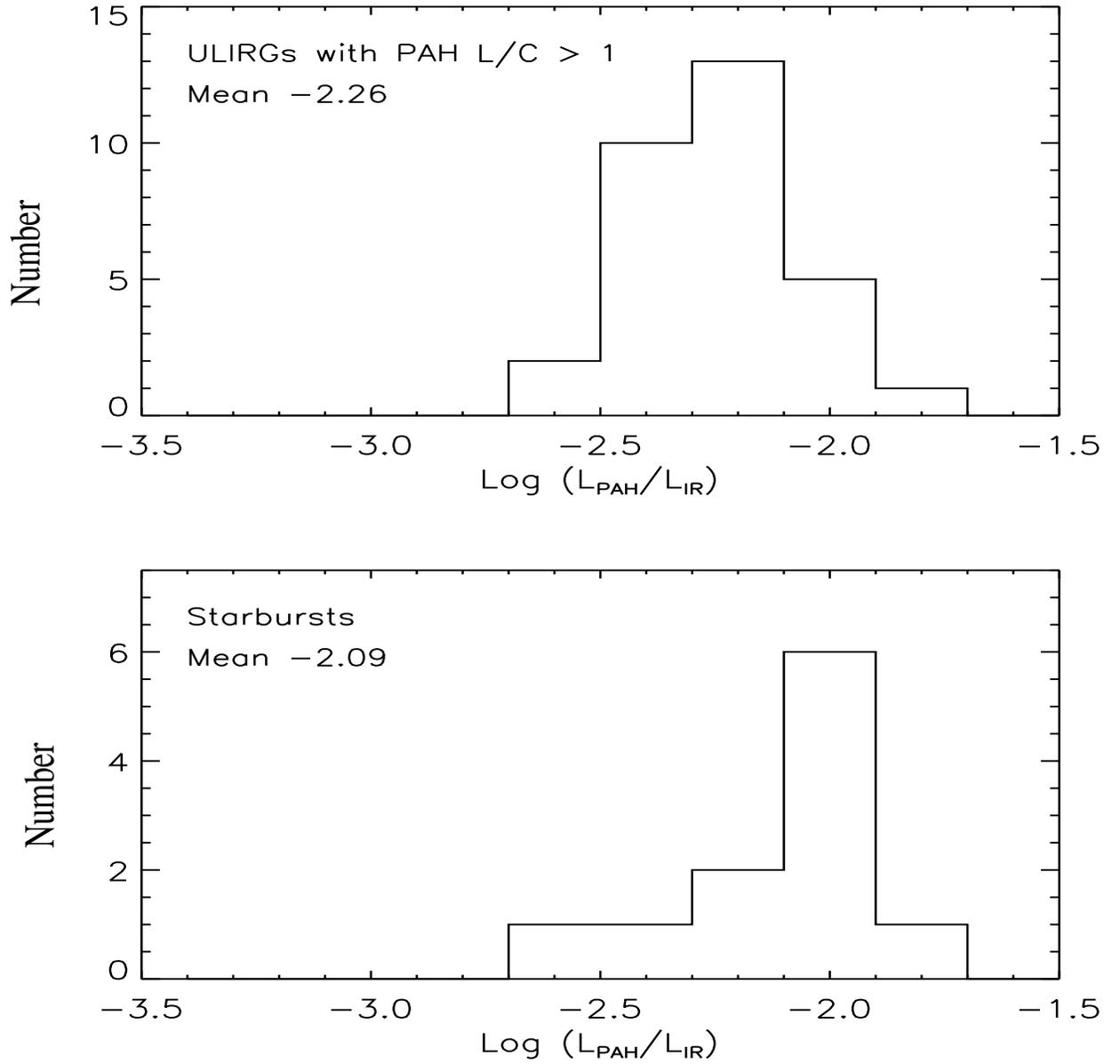}
\caption{Comparison between ULIRG-starbursts and template starbursts}
\end{figure}

\begin{figure}[hbtp]
\figurenum{6}
\psfig{file=rigopoulou6.eps,height=16.92cm,width=18cm}
\caption{extincted M82 spectrum}
\end{figure}

\begin{figure}[hbtp]
\figurenum{7}
\psfig{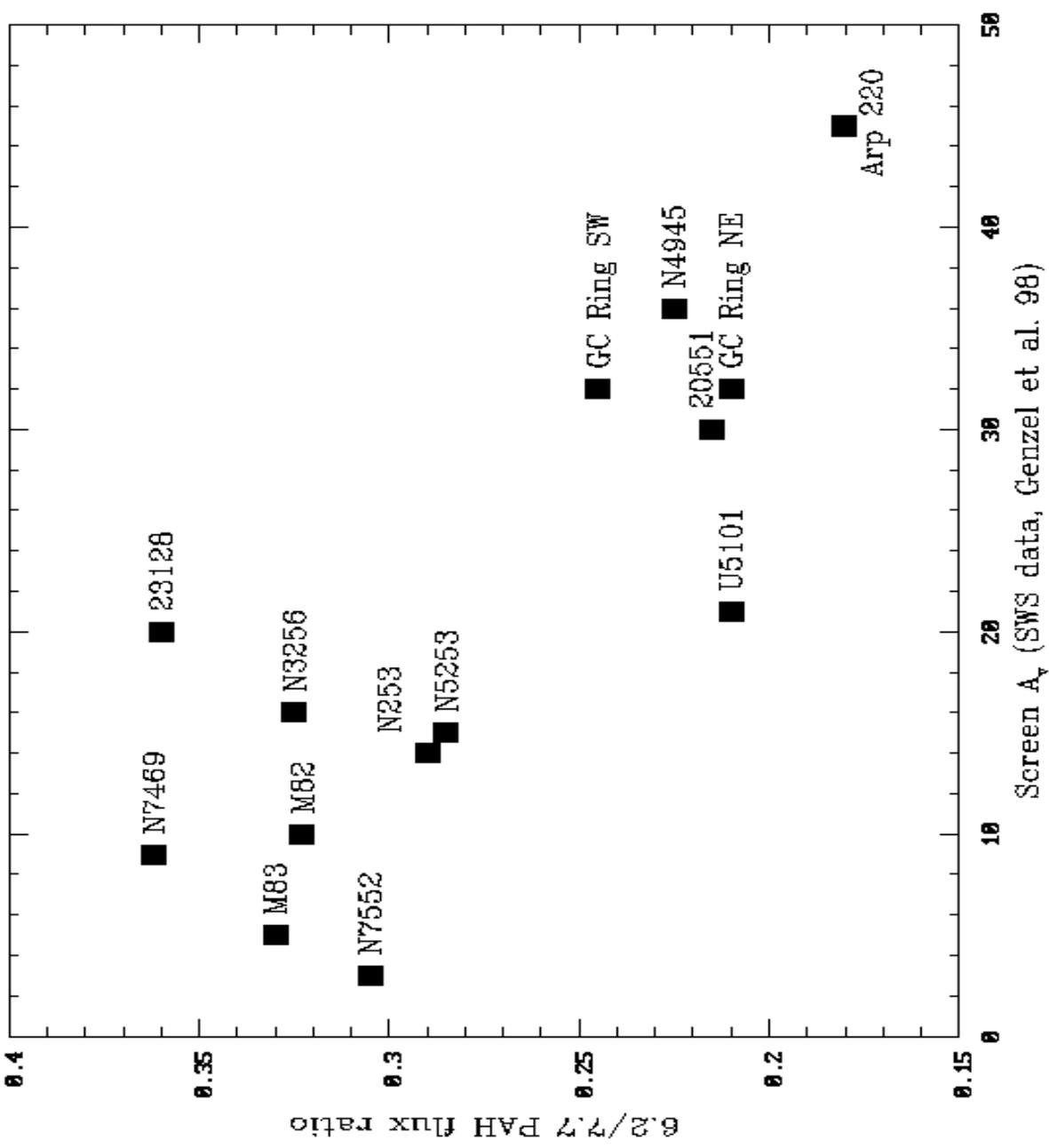}
\caption{Effect of extinction on PAHs}
\end{figure}

\clearpage

\begin{figure}[hbtp]
\figurenum{8}
\psfig{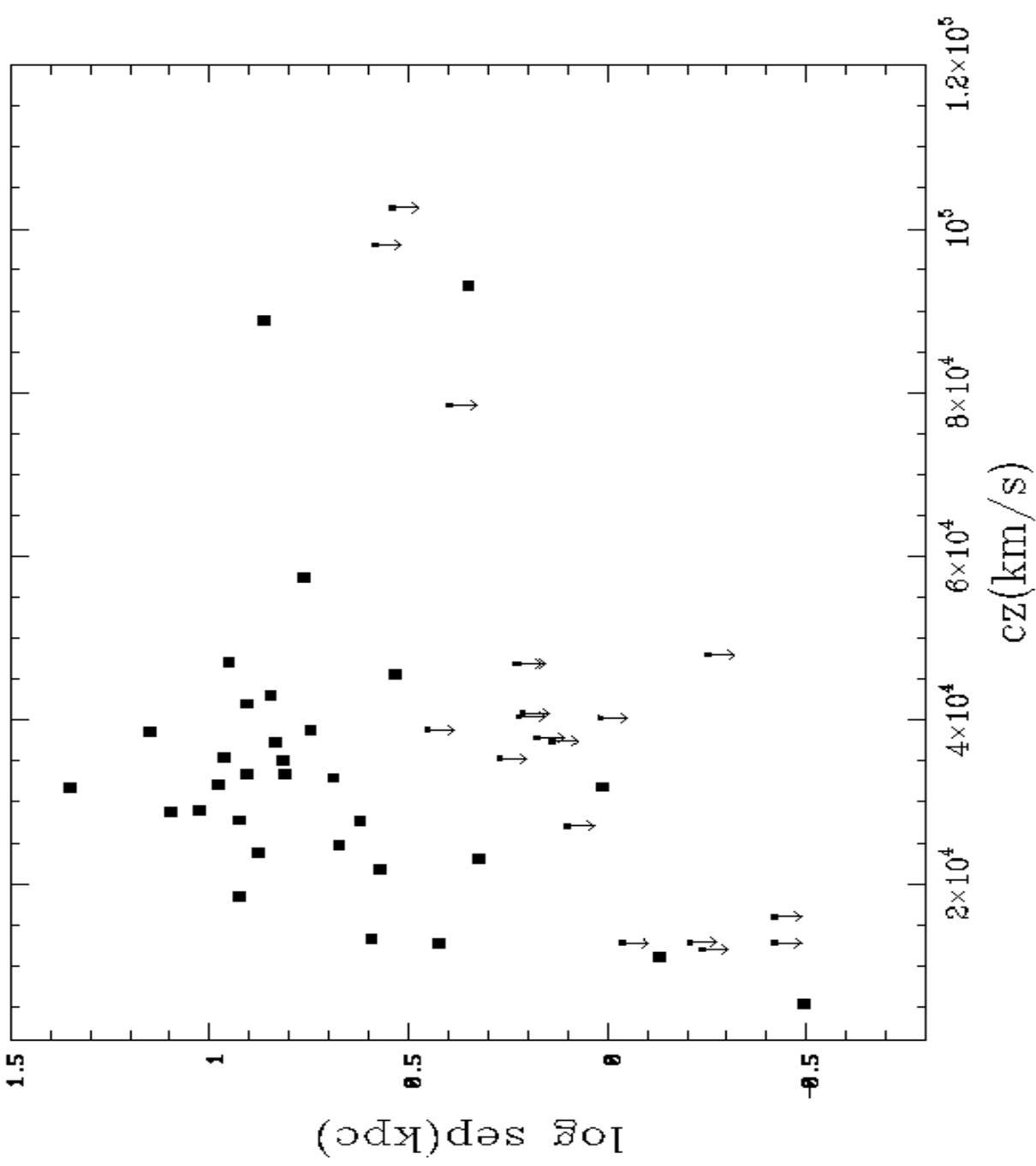}
\caption{Projected nuclear separation as a function of redshift}
\end{figure}

\begin{figure}[hbtp]
\figurenum{9}
\psfig{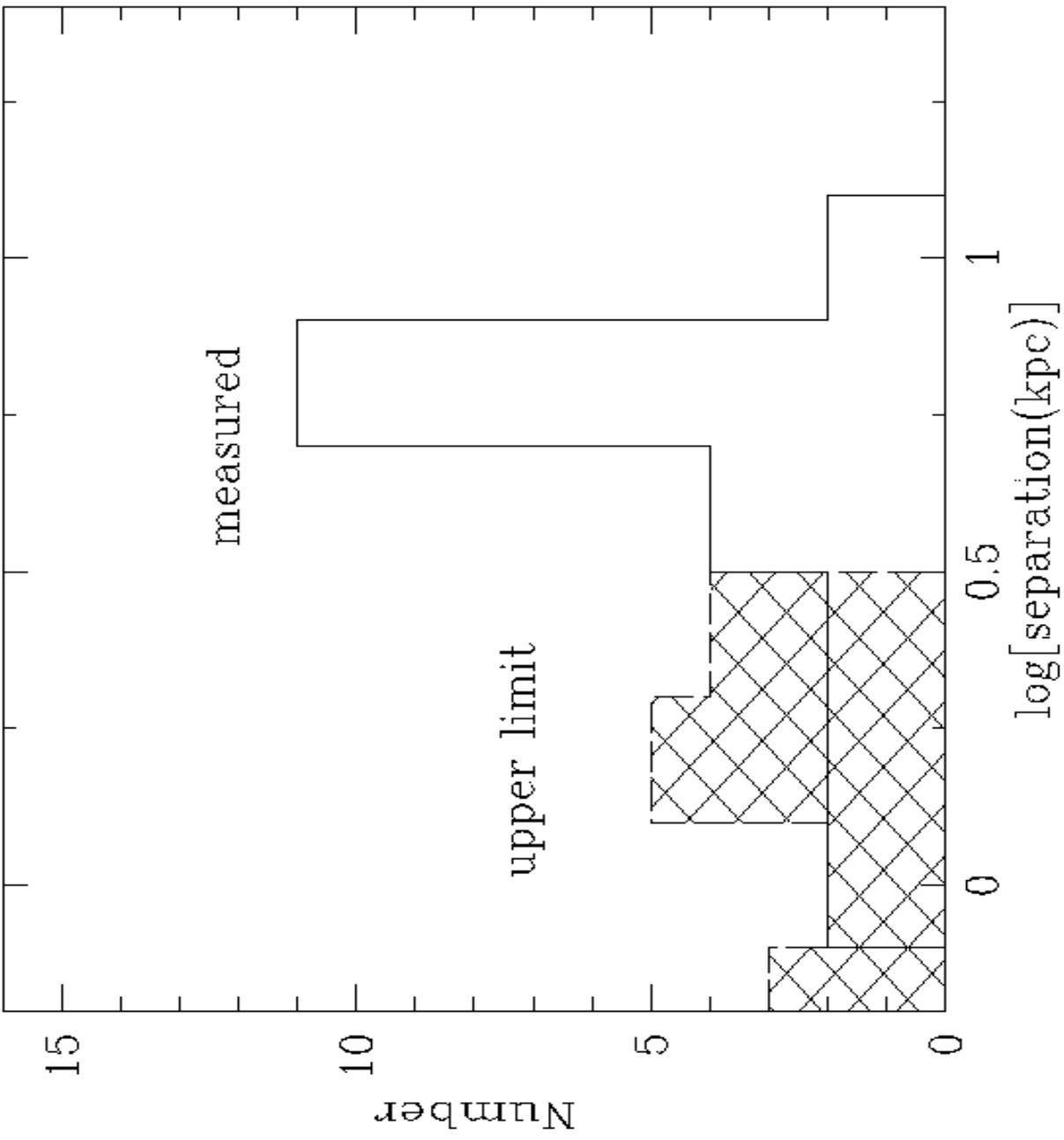}
\caption{Histogram}
\end{figure}

\begin{figure}[hbtp]
\figurenum{10}
\psfig{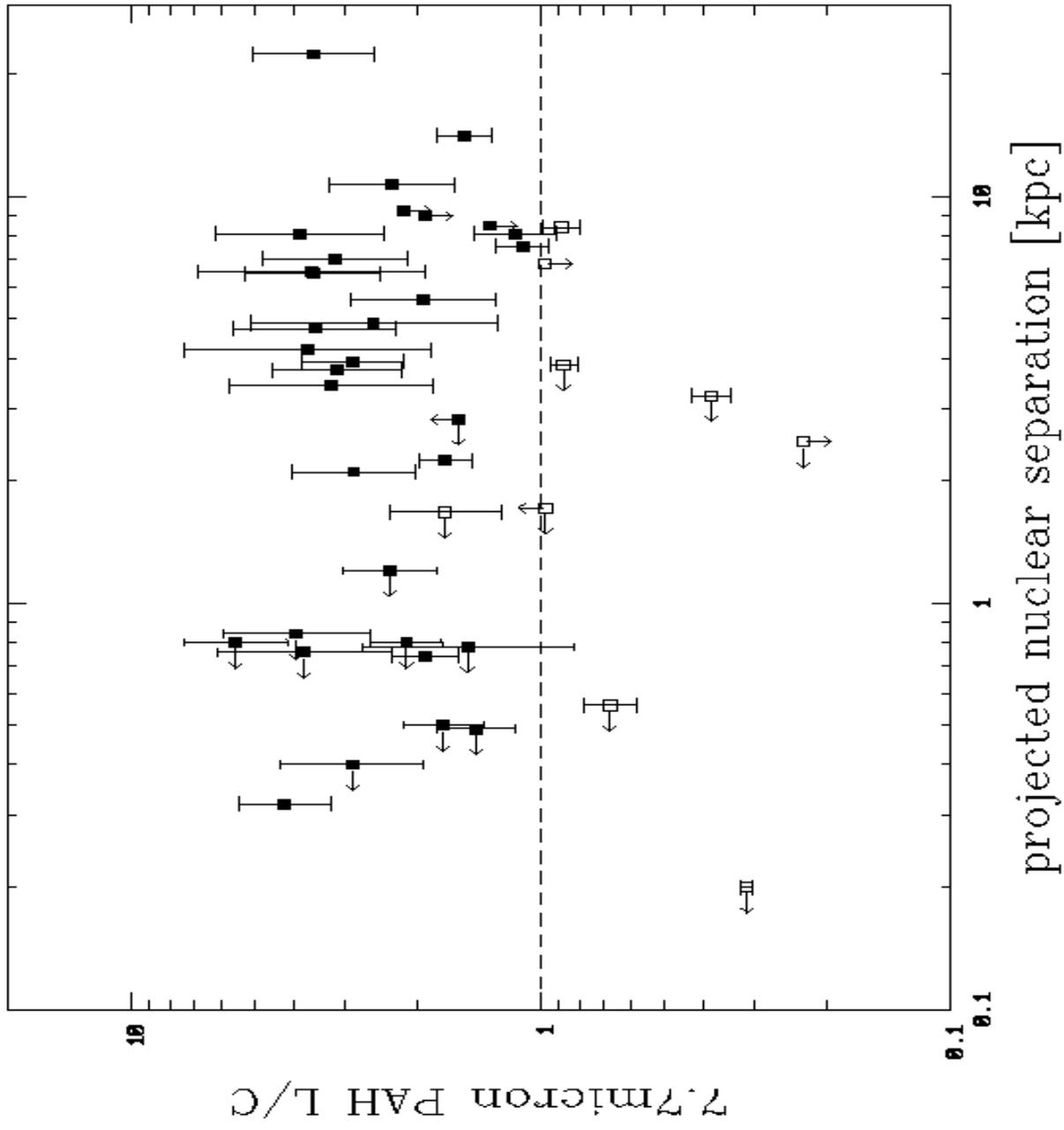}
\caption{Separation plot}
\end{figure}

\begin{figure}[hbtp]
\figurenum{11}
\psfig{file=rigopoulou11.eps,height=16.92cm,width=18cm,angle=-90}
\caption{Luminosity-Separation plot}
\end{figure}

\begin{figure}[hbtp]
\figurenum{12}
\psfig{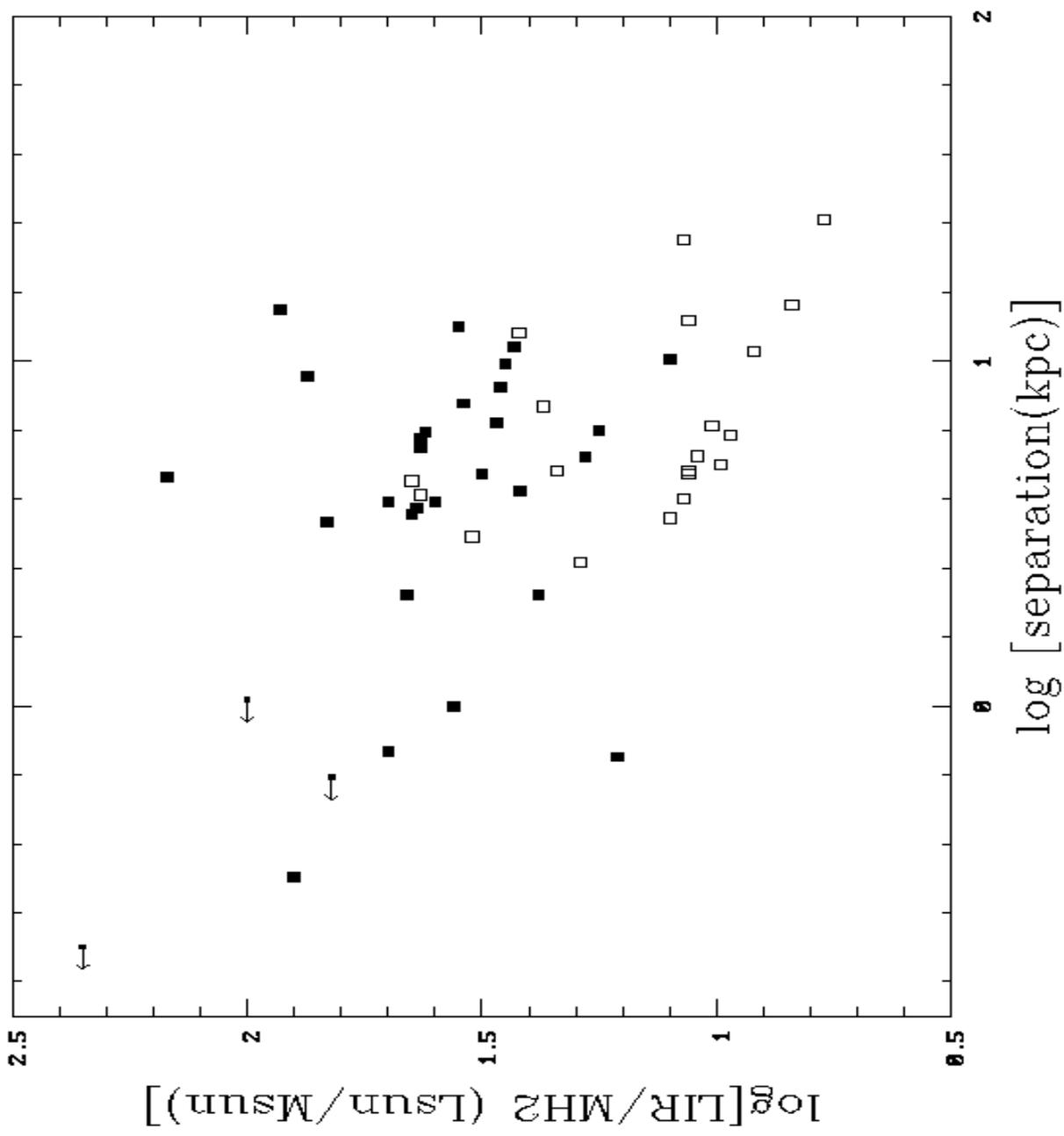}
\caption{Star formation efficiency-Separation plot}
\end{figure}

\begin{figure}[hbtp]
\figurenum{13}
\psfig{file=rigopoulou13.eps,height=16.92cm,width=18cm,angle=-90}
\caption{CO Luminosity-Separation plot}
\end{figure}

\end{document}